\documentclass[a4paper,fleqn]{article}
    \usepackage{natbib}
\usepackage{graphicx}
\usepackage{color}
\newcommand{\cn}{{$C_n^2$}}
\newcommand{\ct}{{$C_T^2$}}

\newcommand{\be}{\begin{equation}}
\newcommand{\ee}{\end{equation}}

\title{Monitoring the optical turbulence in the surface layer at Dome C, Antarctica, with sonic anemometers}

\author{E. Aristidi,$^{1}$\thanks{E-mail: aristidi@unice.fr}
 J. Vernin,$^{1}$ 
 E. Fossat,$^{1}$ 
 F.-X. Schmider,$^{1}$
 T. Travouillon,$^{2}$
 C. Pouzenc,$^{3}$
\\ O. Traull\'e,$^{4}$
 C. Genthon,$^{5}$
 A. Agabi,$^{1}$ 
 E. Bondoux,$^{1}$
 Z. Challita,$^{6}$
 D. M\'ekarnia,$^{1}$
 \\ F. Jeanneaux$^{1}$
 and G. Bouchez$^{7}$
}
\date{\small $^{1}$Laboratoire Lagrange, Universit\'e C\^ote d'Azur, Observatoire de la C\^ote d'Azur, CNRS, Parc Valrose, 06108 Nice Cedex 2, France\\
$^2$ {Thirty Meter Telescope, 1111 South Arroyo Parkway, Pasadena, 91105 CA, USA}\\
$^3${Observatoire Sirene, 84400 Lagarde d'Apt, France}\\
$^4${CNRM-GAME, URA 1357, M\'et\'eo France CNRS, Toulouse, France}\\
$^5${LGGE, UJF-CNRS, 54, rue Moli\`ere BP 96, 38402 Saint-Martin d'H\`eres cedex, France}\\
$^6${Aix Marseille Universit\'e, CNRS, LAM (Laboratoire d'Astrophysique de Marseille) UMR 7326, 13388, Marseille, France}\\
$^7${GEMAC, University of Versailles/CNRS, 45 av. des Etats-Unis, 78035, Versailles Cedex, France}\\
}

\begin{document}
\label{firstpage}
\maketitle

\section*{abstract}
The optical turbulence above Dome C in winter is mainly concentrated in the first tens of meters above the ground. Properties of this so-called surface layer (SL) were investigated during the period 2007-2012 by a set of sonics anemometers placed on a 45 m high tower. We present the results of this long-term monitoring of the refractive index structure constant $C_n^2$ within the SL, and confirm its thickness of 35m. We give statistics of the contribution of the SL to the seeing and coherence time. We also investigate properties of large scale structure functions of the temperature and show evidence of a second inertial zone at kilometric spatial scales.

\section{Introduction}
During the last two { decades}, the Antarctic Plateau has been attracting interest from the astronomical community. Since the first site-testing experiments at the South Pole (\cite{1999AAS..134..161M} a lot of work has been done, especially after the advent of the French-Italian station Concordia at the summit of Dome C. It revealed very interesting qualities for astronomy. Its altitude (3300m) { combined with} low humidity are expected to give excellent transparency and low sky background emissivity in the infrared (\cite{2004PASP..116..482L}. The good meteorological conditions  { combined with its} location near the Pole (75$^\circ$S) gives access to high temporal coverage (\cite{2007PASP..119..127M}. In winter the seeing was found to be excellent (about 0.4~arcsec) above a thin surface layer (SL) of thickness close to 30m (\cite{2004Natur.431..278L, 2008PASP..120..203T, 2009AA...499..955A}. Seeing stability studied by (\cite{2010AA...517A..69F} reported periods of 7-8~hours where the seeing was continuously below 0.5 arcsec.

Other polar stations were investigated in Antarctica:  Dome A (\cite{2010PASP..122.1122B}, and Dome F (\cite{2013AA...554L...5O}. Similar properties were found at these locations, in particular this thin SL which gives access to the free atmosphere seeing at elevations of a few tens of meters above the ground. This is also supported by numerical simulations by (\cite{2006PASP..118.1190S, 2008MNRAS.387.1499H, 2009MNRAS.398.1093L, 2010MNRAS.403.1714L, 2011MNRAS.411..693L}. Therefore the characterization of the SL, both in terms of intensity and vertical structure, is  critical to the determination of methods to compensate its effects (\cite{2009PASP..121..668T, 2010EAS....40..157C}. So far, our knowledge of the SL comes from several instruments, but  each one have specific flaws. Early SODAR measurements, for example, didn't have the vertical resolution to resolve the SL turbulence (\cite{2004Natur.431..278L}. Balloon-borne microthermal sondes, do have the vertical resolution but lack the temporal resolution necessary to obtain a statistically meaningful data set (\cite{2008PASP..120..203T}. Recent SODAR monitoring do have temporal and spatial resolution, but do not provide a reliable estimate of the turbulence intensity (\cite{2014AA...568A..44P}.

\begin{figure}
\includegraphics[width=8cm]{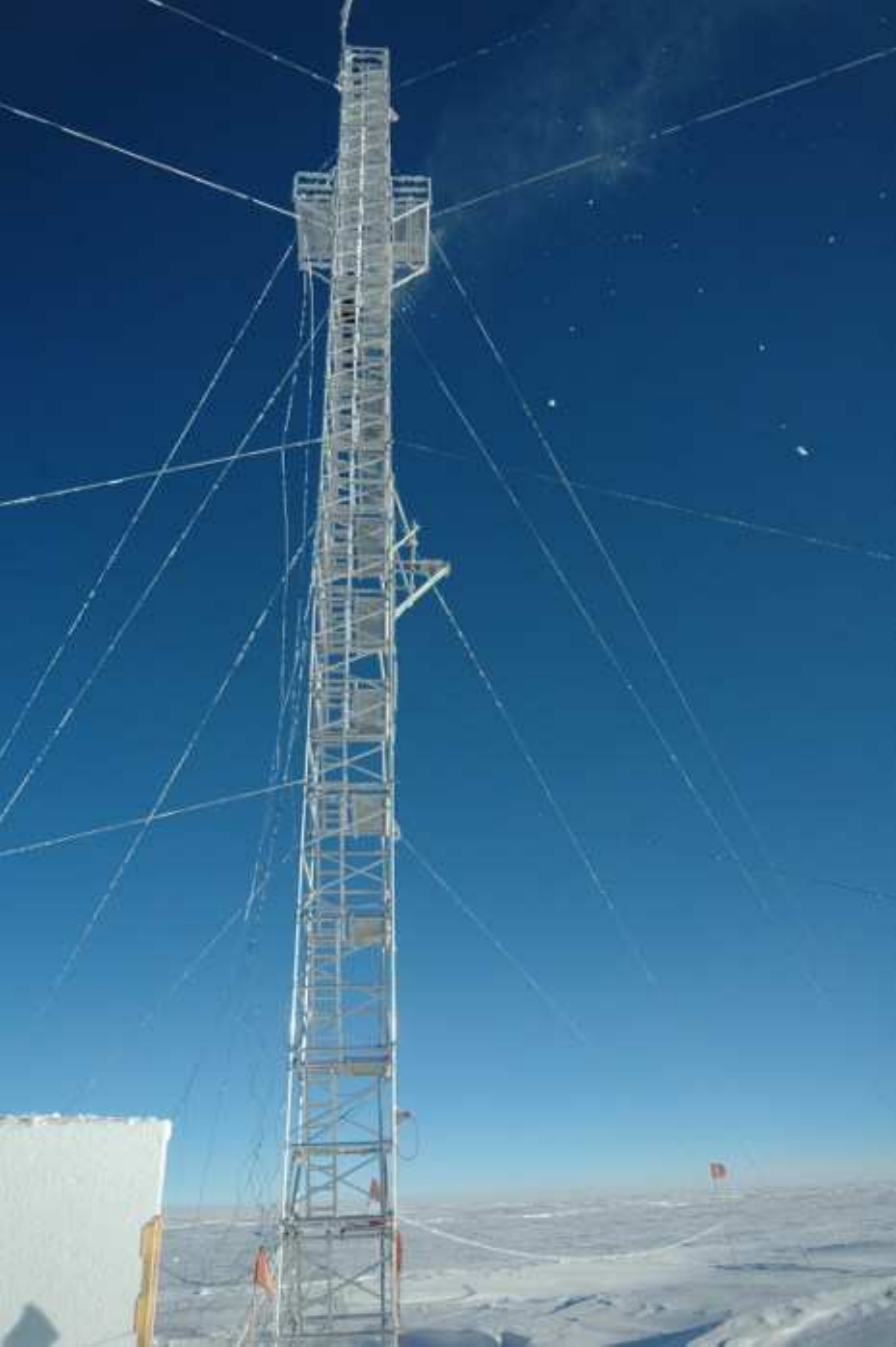}
\caption{The 45m tower at Dome C. Computers and electronics are located in the shelter at the foot of the tower.}
\label{fig:tower}
\end{figure}

In situ measurements therefore became a viable option. The presence of a 45m tower (Fig.~\ref{fig:tower}) allowed the possibility of these measurements. First attempts were made during the first winterover in 2005, with microthermals sensors (\cite{2005PASP..117..536A}   at different locations on the tower. However  they appeared to be too fragile in the windy and icy conditions and no reliable results could be obtained. An alternative to microthermal sensors which uses similar principles to measure optical turbulence without the issue of brittleness is sonic-anemometers. Their use in the study of atmospheric turbulence was  discussed by (\cite{1979dmuf.proc..551K} and more recently by (\cite{2015JPhCS.595a2036T}. Their principle is to make a high speed measurement of temperature and wind velocity vector, which are processed to obtain an estimate of the optical turbulence structure constant  $C_n^2$. Using several sonic-anemometers along the height of the tower allows to make discrete measurements of the turbulence profiles only limited in vertical resolution by the number of instruments used. Our instrumentation and the first result were introduced in (\cite{2008SPIE.7012E.147T, 2010EAS....40..115T}. The scope of this paper is to present our data processing method based upon temperature structure functions, and local turbulence measurements in terms of refractive index structure constant \cn\ obtained between 2007 and 2012.

The paper is organised as follows. Section~\ref{par:theory} recalls  theoretical concepts about the optical turbulence. The experimental setup is described in section~\ref{par:instrument}. Data processing, error analysis, sources of biases and instrument calibration are presented in section~\ref{par:dataprocessing}. Results are shown in section~\ref{par:results}.

\section{Theory}
\label{par:theory}
Turbulence discussed in this article is the optical turbulence, which is related to spatial and temporal heterogeneities of the refractive index. These heterogeneities results from both the meteorological turbulence (vortices themselves) and gradients of the temperature and the wind speed (in particular the vertical temperature gradient). At Dome C, during the polar winter, strong temperature inversions are observed in the first tens of meters above the snow  surface~(\cite{2008PASP..120..203T, 2013JGRD..118.3218G}. This is a very stable situation and  meteorological turbulence is mostly inhibited in this case. But the few {meteorological turbulence} that remains induces strong optical turbulence because of these strong temperature gradients.

Atmospheric optical turbulence is often described by a set of quantities such as the seeing $\epsilon$, the isoplanatic angle $\theta_0$, the coherence time $\tau_0$ or the spatial coherence outer scale ${\cal L}_0$. These parameters all result from an integration along the line of sight of local quantities such as the wind speed and the structure constant of the refractive index \cn. Theoretical basis of the effects of optical turbulence in astronomy are described in the review paper of (\cite{1981PrOpt..19..281R}, and we will here pick up some particular points of interest for the present study.

Turbulent air flows are characterized by random wind speed and temperature fluctuations. The  temperature structure function is among the statistical quantities describing the temperature fluctuations at two points separated by a distance $\rho$ and is given by 
\be
D_T({\bf \rho})=\langle (T({\bf r})-T({\bf r+\boldmath\rho}))^2 \rangle
\ee
where $r$ is the position of one point, $T$ the temperature and $\langle \rangle$ stands for ensemble average. This function behaves as a power law in $\rho^\frac{2}{3}$ when $\rho$ is comprised between the dynamic inner scale $l_0$ and the outer scale $L_0$. These scales can be seen as the sizes of the smallests and largests eddies of the turbulent air flow. Outside the interval $[l_0,L_0]$ (so-called inertial range) the structure function tends towards a constant. Within the inertial range, one defines the structure constant of the temperature $C_T^2$ as
\be
D_T({\bf \rho})=C_T^2\, \rho^\frac{2}{3}
\ee
and making use of the Gladstone's law which links the refraction index to the temperature, the structure constant of the refractive index \cn\ expresses as
\be
\label{eq:ctcn} 
C_n^2=6.24\, 10^{-9}\ C_T^2\, P^2 \, T^{-4}
\ee
with $T$ in Kelvin and the pressure $P$ in hPa. The seeing as seen by a telescope at an elevation $h_0$ is computed by the following integral over the altitude $z$ ($\epsilon$ is here in radian, and $\lambda$ is the wavelength): 
\be
\epsilon=5.25\, \lambda^{-1/5} \, \left[ \int_{h_0}^\infty C_n^2(z)\, dz \right]^{\frac{3}{5}}
\label{eq:cneps} 
\ee
The coherence time is deduced both from the \cn\ profile and the wind speed profile $|V(z)|$
\be
\tau_0=0.058\, \lambda^{6/5} \, \left[ \int_{h_0}^\infty |V(z)|^{5/3}\, C_n^2(z)\, dz \right]^{-\frac{3}{5}}
\label{eq:cntau0} 
\ee

\section{The sonic anemometers}
\label{par:instrument}
The experiment is based upon a set of sonic anemometers installed on a tower located about 900~m West from the Concordia buildings. These anemometers are modified Sx probes from Applied Technnologies Inc. They are composed of 3 pairs of ultrasonic transducers capable of measuring the wind speed in three orthogonal axes (namely $U$, $V$ and $W$ where $W$ is the vertical axis), as well as the temperature $T$. They are placed on horizontal bars pointing in the upwind direction to void the possibility of local effect due to the mast. Principle of the measurement involves the dependance of the transit time of acoustic pulses with the velocity of the wind. The time difference between two pulses travelling between a pair of transducers in opposite directions is indeed proportionnal to the wind speed, whereas the sum of these transit times gives the speed of sound, related to the temperature (see (\cite{1976JApMe..15..607F} and references therein). The four numbers are provided at a rate of 10 Hz, a sampling high enough to infer the properties of turbulence, and in particular to estimate the structure constants $C_T^2$ and \cn\ as described in \S\ref{par:cnsonic}. Technical specifications of the anemometers are given in Table~\ref{tab:sonics}

These instruments were modified in order to operate in the low temperatures of the Antarctic Plateau. The modification consists of a layer of aerogel which thermally insulates the sensing parts of the instruments, as well as a wrap of heating resistances that warms the units and protects them from ice formation. A cycle alternates between heating the probes and measuring (no measurement is possible during the heating since it would create unwanted local turbulence). After some trials we found that a 20~mn period cycle with 10~mn of heating and 10~mn of measurements is a good compromise. {The first few minutes (typically 2-3mn) after the heater has been turned off are contaminated by noise and are eliminated during the data processing}. Still, in winter it is sometimes necessary to climb on the tower and manually remove the accumulated snow.

Three of these anemometers were installed in November 2006 (the tower was then 30~m high) at heights 8, 17 and 28~m above the snow surface. Because of sinking and snow accumulation, heights tend to decrease at a rate of $\simeq$ 10cm/year. These three sonics were operated during the whole first year, though the 17m one suffered from a technical problem that resetted its calibration parameters. The probe was recalibrated in January 2008 and the data could be corrected afterwards by software. The tower was extended by 15m in December 2007, to attain a height close to 45m, and three additionnal anemometers were installed on the new section. The heights of the sonics became 8m, 17m, 24m, 31m, 39m and 45m (the heights of the instruments were remeasured at the end of 2010 by (\cite{2013JGRD..118.3218G} who found values a few \% lower). This 6-sonic setup was first operated in January 2008. The 2008 data present some gaps due to various problems (crash of the computers, problems with heaters power supplies). Then in 2009 we obtained a very good set of data, though the 24m probe stopped to work in March. It was replaced in November 2009 by the 45m one, and {the experiment ran with} 5 sonics in 2010. Two other anemometers broke down at the beginning of 2011, as well as the entire heating system which can not be repaired in winter due to the harsh conditions. 3 sonics could be repaired for the winter 2012, and the system worked well up to May 2012.

\begin{figure}
\includegraphics[width=8cm]{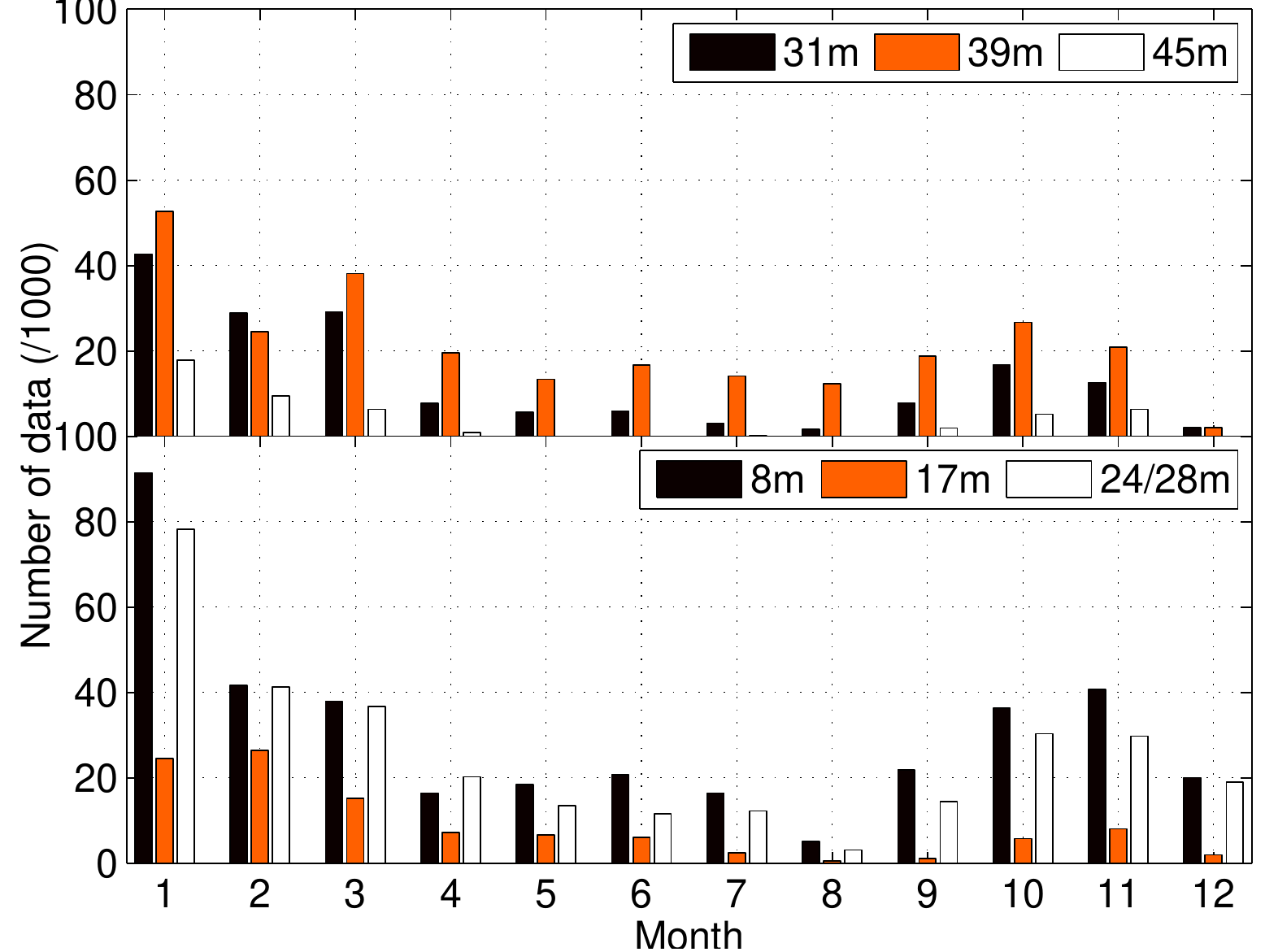}
\caption{Number of data samples collected by each sonic every month from November 2006 to May 2012. One data sample corresponds to one minute of measurement. Top: the three upper sonics. Bottom: the three lower ones (the third was at height 28m in 2006-2007, then at height 24m).}
\label{fig:ndata}
\end{figure}

The journal of the observations is summarized in Fig.~\ref{fig:ndata} showing the number of data collected month by month by each sonic, after filtering of the bad points. It shows, unsurprisingly, that there are much more data in summer (ground temperature around -30$^\circ$C) than in winter (temperature around -65$^\circ$C). Il also shows that the highest anemometer (45m) collected less data than lower ones. {Indeed the temperature and the humidity of the air increase with the altitude (\cite{2013JGRD..118.3218G}. The metallic structure of the upper part of the tower and of the highest anemometers are colder than the air surrounding and tend to ice up more frequently.}

\begin{table}
\begin{center}
\begin{tabular}{|l|l|}\hline
Temperature range & -80$^\circ$C to +60$^\circ$C \\
Wind speed range & 0 to 30 m/s\\
Wind speed accuracy & 0.03 m/s\\
Temperature accuracy & \\
\ \ \ - Absolute & 2$^\circ$C \\
\ \ \ - Relative & 0.1$^\circ$C\\ \hline
\end{tabular}
\end{center}
\caption{Technical specifications of the sonic anemometers (from App. Tech. Inc).}
\label{tab:sonics}
\end{table}

\section{Data processing}
\label{par:dataprocessing}
\subsection{\cn\ measurements from sonic data}
\label{par:cnsonic}
Sonic anemometers are connected to a data packer formatting the output of all the units so that they can be read directly from the RS232 port of a PC. One obtains an ascii file with 4 columns by sonic (containing the velocities $U$, $V$, $W$ and the temperature $T$) and one line every 0.1 second. That represents a volume of data of about 50 MB/day. A first preprocessing program scan these raw data in order to remove outliers.

Deriving the structure constant \cn\ implies the estimation of the structure function $D_T^2(\rho)$, i.e. to estimate the temperature fluctuations at two points separated by the distance $\rho$. As we have only one sensor at a given altitude, we use the Taylor's hypothesis to transform $\rho$ in $\bar v\, \Delta t$ ($\bar v$ being the mean wind speed and $\Delta t$ a time interval between two measurements) and estimate the structure function as the temporal average
\be
D_T^2(v\, \Delta t)=\langle (T(t)-T(t+\Delta t))^2 \rangle
\ee
Two methods were compared to compute the structure constant $C_T^2$, they give similar results. 
\subsubsection{First method: structure function}
\label{par:fs}
The aim is to estimate the full function $D_T^2(\bar v\, \Delta t)$ by taking several values for $\Delta t$ (the sampling period is 0.1~second, so all the $\Delta t$ are multiples of 0.1 second). The temporal average has to be calculated on a time interval $\tau$ long enough to ensure statistical significance, but shorter than the characteristic time of evolution of $C_T^2$. After some trials we found $\tau=30$~mn as a good compromise. And the wind speed modulus $\bar v$ which intervenes in $\bar v\, \Delta t$ was taken as a sliding average of $v=(U^2+V^2+W^2)^{1/2}$ over one second.

\begin{figure}
\includegraphics[width=8cm]{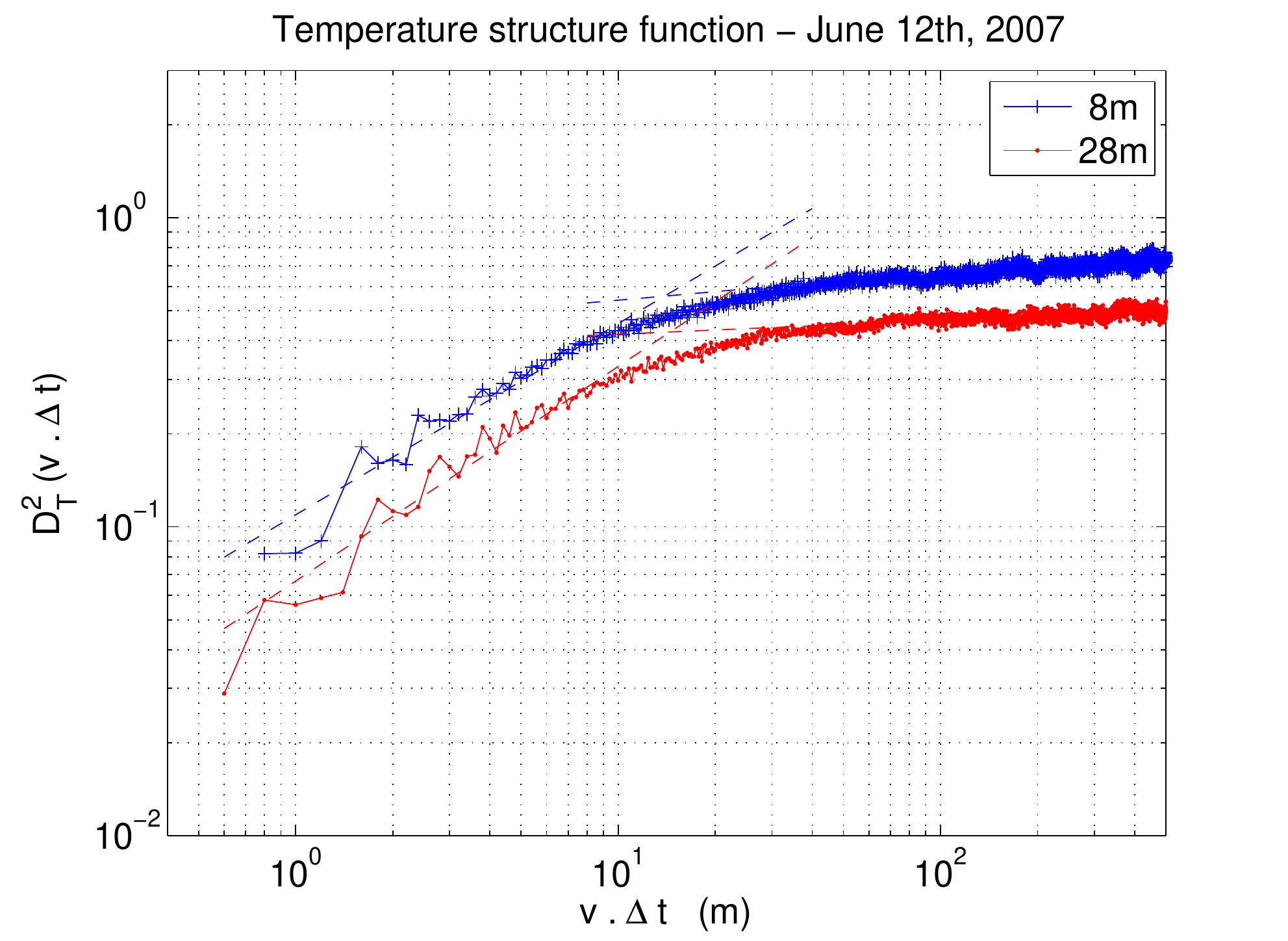}
\caption{Structure functions computed from data taken on June 12th, 2007. The upper curve corresponds to the 8m sonic, the lower one to the 28m sonic. The dashed lines are least-square fits of the {inertial} domain and on the saturation domain (large values of $\bar v \Delta t$).}
\label{fig:fstruct}
\end{figure}

Fig. \ref{fig:fstruct} shows an example of structure functions computed on data taken on June 12th, 2007. The two curves correspond to the lowest (8m) sonic, and the 28m one. In log-log scale, both curves display a linear part in the interval [1m, 10m]. A least-square fit of a function $C_T^2\, \rho^\alpha$ (with $\rho=\bar v\ \Delta t$) was performed in this interval, and is displayed on the graph as a dashed line. It gave ($C_T^2=0.11\pm 0.04$ m$^{-2/3}$, $\alpha=0.6\pm 0.2$) for the 8m sonic, and ($C_T^2=0.07 \pm 0.02$ m$^{-2/3}$, $\alpha=0.7\pm 0.2$) for the 28m sonic. The expected value of $\alpha$ being 2/3 in the inertial domain $[l_0,L_0]$. For large values of $\rho$ the structure function saturates as expected within the saturation region. 
{
Slopes were computed by least square fit in the saturation region, i.e. in the interval $\rho\in [100, 200]$m. This interval was selected to make sure that it is well above the outer scale, and it has enough data points to reduce statistical noise. In the example of fig.~\ref{fig:fstruct} we found slopes $\alpha=0.09$ for the 8m sonic and $\alpha=0.04$ for the 28m one}. An estimation of the outer scale can be obtained at the intersection of the fits {inside the inertial zone and inside the saturation region}. We obtained $L_0=14$m for the two heights.

\begin{figure}
\includegraphics[width=8cm]{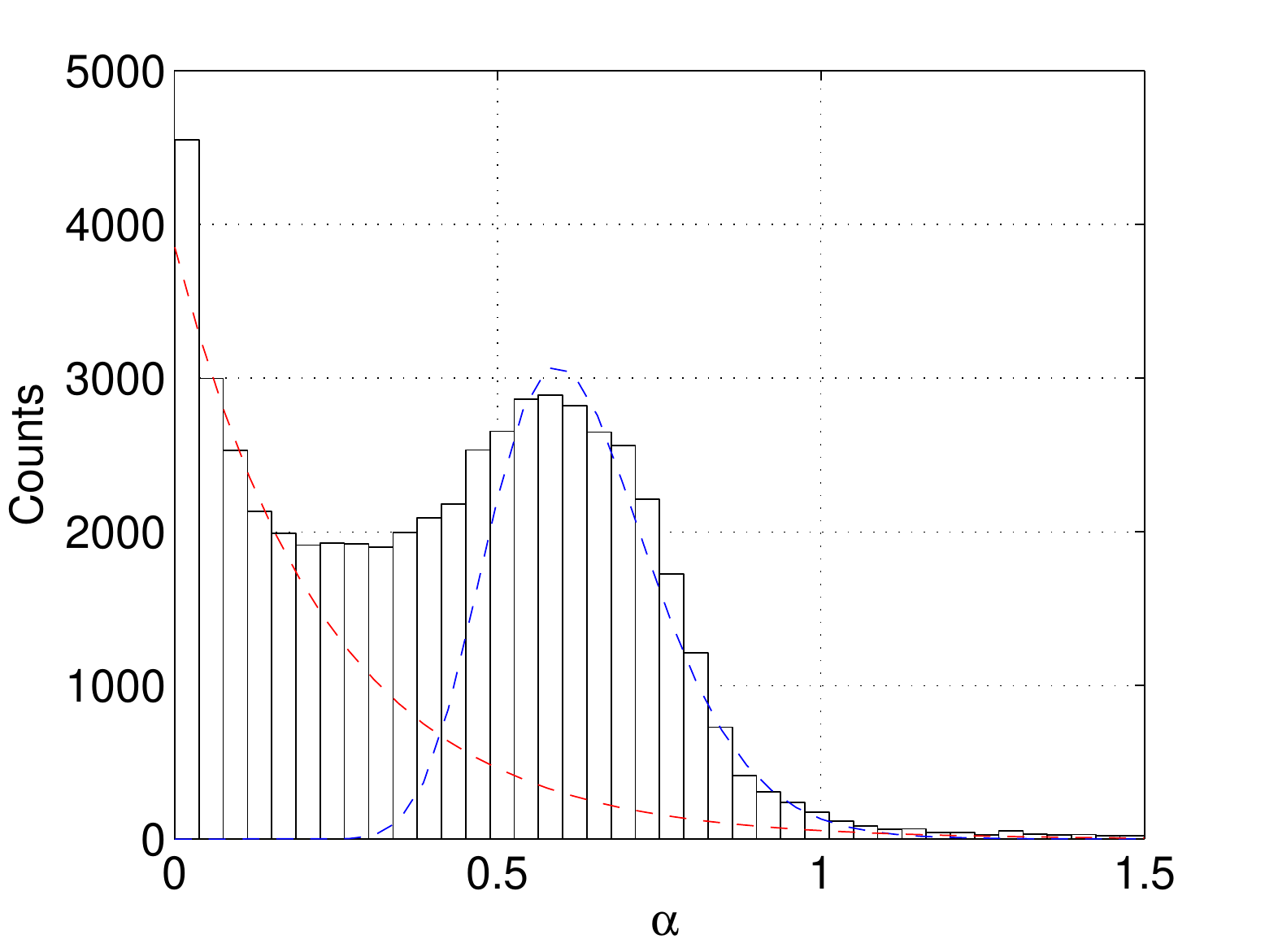}
\caption{Histogram of the slopes $\alpha$ of the structure functions $D_T(\rho)$ of the sonic temperatures in the interval $\rho\in [0.6\mbox{m},4]\mbox{m}$. This histogram was computed from all 6 sonics data collected during the years 2008-2011. The dashed lines are (i) a negative exponential fit of the first part of the histogram for small values of $\alpha$, and (ii) a log-normal fit of the bump centered at $\alpha=0.6$.}
\label{fig:pentefs}
\end{figure}

Structure functions were computed on the whole set of data, and for each function a least-square fit, as described in the example above, was performed on the interval $\rho\in[0.6\mbox{m},4\mbox{m}]$. { This interval was selected after some trials, but its boundaries have little effect on the computed slopes and \ct : the change is 3\% on the slope and 2\% on \ct\ if we use $\rho\in[1.2\mbox{m},4\mbox{m}]$, this is well below the error bars.} Histogram of obtained slopes is shown in Fig.~\ref{fig:pentefs}. It displays two features: a decreasing exponential-like curve for very small value of the slope $\alpha$, and a hump centered on the value $\alpha=0.6$. This histogram mixes the data from all the anemometers, but individual ones look all the same, with even a more distinct separation beween the two structures. Also there does not seem to be a dependance with the season (excepted that there are less data in winter so the histograms are more noisy). These two structures are characteristic of two different situations. The negative exponential is observed for very small slopes (its with is $\Delta\alpha=0.15$) which indicates that corresponding data are in the saturation domain of the structure function and cannot be used to compute the parameter $C_T^2$. 
{Indeed various types of data correspond to this case: frost or snow on the sensors (causing a random shape of the structure function), contamination by the heaters (the damping time after the heaters have been turned off is a few minutes), and other technical problems. These data will subsequently be removed from the analysis}.
The bump corresponds to slopes around $\alpha=2/3$, with a dispersion $\Delta\alpha\simeq \pm 0.2$ (FWHM of the fitted log-normal curve is 0.4) of the same order of magnitude than the uncertainty given by the least-square fit of the structure function. Data corresponding to this bump are in the inertial domain of the structure function and $C_T^2$ can be derived from them. They represent about 40\% to 60\% of the of data collected by each sonic.

\subsubsection{Second method: direct calculation}
\begin{figure}
\includegraphics[width=8cm]{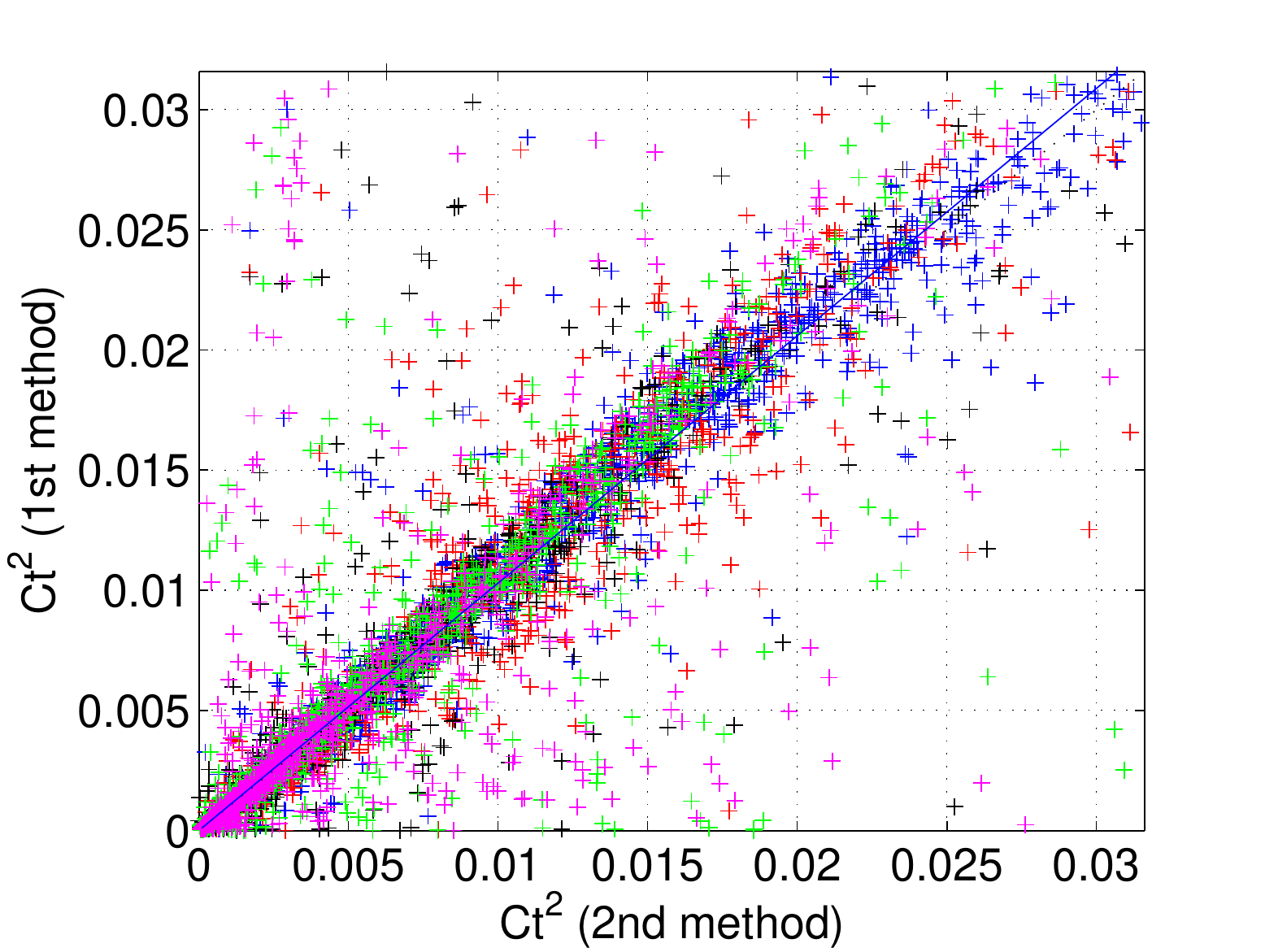}
\caption{Comparison of $C_T^2$ values computed from the first method (fit of the inertial domain of the structure function) and the second method (direct computation from consecutive temperature values). Colors correspond to sonic elevation with the same convention as Fig.~\ref{fig:cn2month}. The straight line is a least-square fit of the data, its slope is 1.02.}
\label{fig:comparct2}
\end{figure}

This analysis {presented in the previous section} appears to be very efficient and provides a way of sorting the useful data, but the computation is very time consuming and one obtains only one value of $C_T^2$ every 30~mn. An alternative is to compute one single point of the structure function for the smallest possible value of the time interval $\delta t=0.1$s, i.e. the quantity $[T(t)-T(t+\delta t)]^2$. The structure constant $C_T^2$ is estimated by the quantity
\be
C_T^2=\left \langle \frac{[T(t)-T(t+\delta t)]^2}{[v(t) \delta t]^{2/3}} \right \rangle_\tau
\label{eq:calcct}
\ee
where the temporal average is made on a time $\tau=1$mn. This method is a shortcut allowing quick computation of one value of $C_T^2$ every minute.  Fig.~\ref{fig:comparct2} shows a comparison between $C_T^2$ computed both ways. The curve was obtained as follows: structure functions were computed every 30~mn, {giving} a first set of $C_T^2$ values (we selected only data for which the slope of the structure function  $\alpha>0.4$. For the same 30mn periods, direct computation of 30 values of $C_T^2$ were calculated and averaged. The plot on Fig.~\ref{fig:comparct2} exhibits a strong linear relation between the two sets of $C_T^2$ with a slope of 1.02 and a correlation of 97\%. The two method appears then to give consistent results. {There is some scatter around the straight line of slope 1.02, of about 20\%, which is of the same order of magnitude than the statistical error on \ct\ (see Sect.~\ref{par:errct}).}
\\

Finally we adopted the following algorithm for deriving \cn\ from sonic data:
\begin{itemize}
\item Split the data into 30mn intervals
\item Compute a structure function and check the slope $\alpha$.
\item If {$\alpha<0.4$} reject the 30mn interval
\item Otherwise compute 30 values of $C_T^2$ by the second method.
\item Calculate \cn\ with Eq.~\ref{eq:ctcn}, taking for $T$ the average temperature over one minute, and for $P$ the mean yearly pressure of the site (645~hPa) since pressure fluctuations at Dome C are only a few percent. 
\end{itemize}
Hence the structure functions were used only as a data filter, and the $C_T^2$ coefficients were computed via the direct method.

\subsection{Dynamic outer scale $L_0$}
\label{par:L0}
\begin{figure}
\includegraphics[width=8cm]{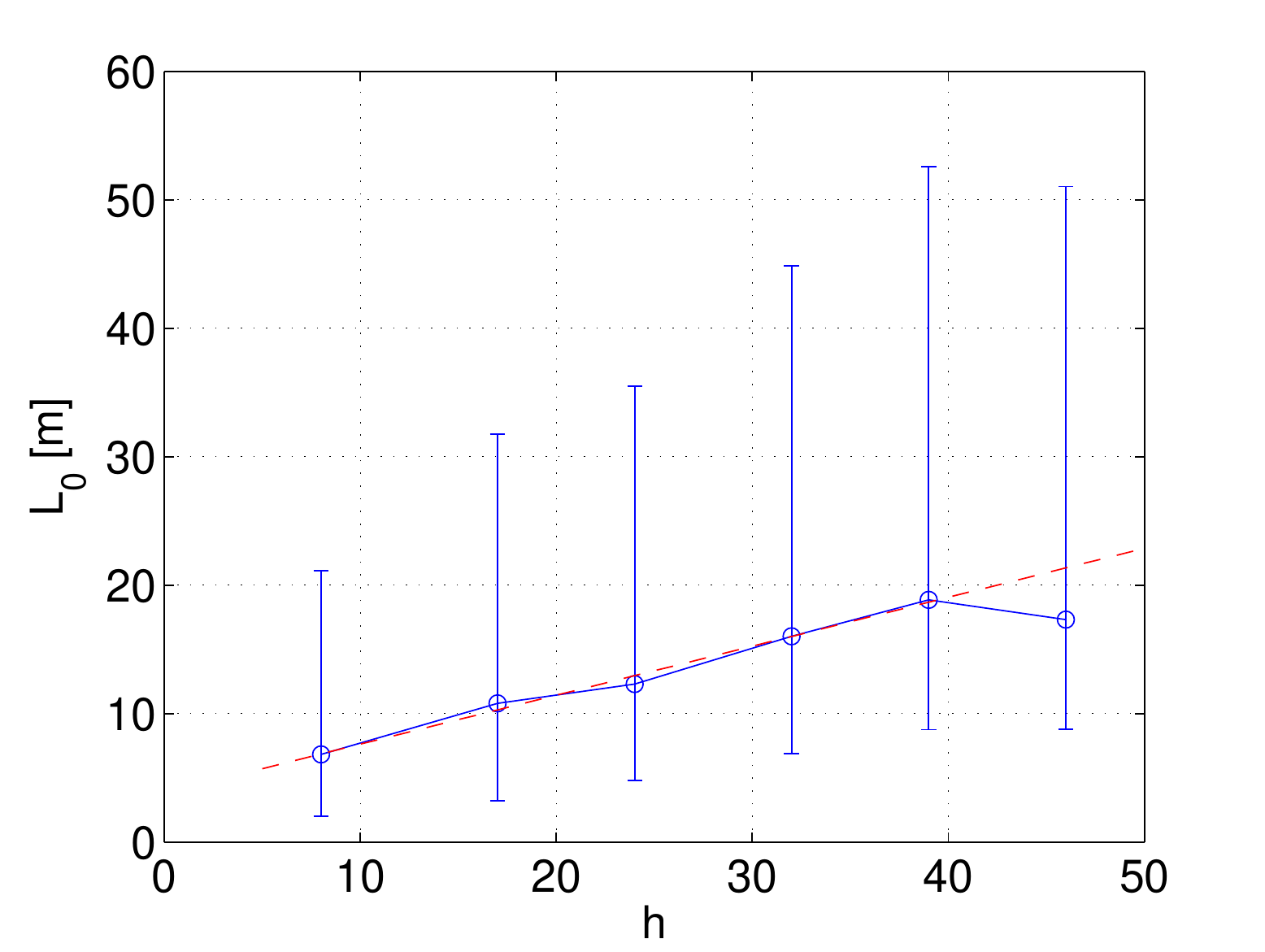}
\caption{Outer scale $L_0$ as a function of the height of the anemometer. Error bars are the intervals containing 50\% of the data. The dashed line is a least-square fit obtained on the five first points.}
\label{fig:L0vsh}
\end{figure}

As stated in section~\ref{par:fs}, the computation of the structure function allows to calculate the outer scale $L_0$ from a set of several tens of minutes of sonic data. Fig.~\ref{fig:L0vsh} shows the graph of the median value of $L_0$ (computed from 1~hour data samples) as a function of the height of the anemometer. The data were filtered so that the slope $\alpha$ of the log of the structure function in the inertial domain is greater than 0.4. Several thousands of reliable values for $L_0$ were obtained for each altitude.

The five first points of the graph appear to be almost aligned; a least-square fit of a first order polynom on these points, {weighted by the inverse square of the error bars} shows a slope of $0.38\pm 0.04$. It is indeed well-known that the size $L$ of the eddies near the ground is $L=0.4 h$ where $h$ is the altitute (see for exemple Nakayama and Boucher, 2000). {Despite large error bars,} this agreement gives us confidence in the validity of our outer scales and consolidates our data processing method.

The last point of the graph, corresponding to the anemometer at height 45m, is the less significative (2000 values instead of 4000 for the others).

\subsection{Seeing and coherence time in the surface layer}
The seeing is derived by integrating the structure constant profile $C_n^2(z)$ over the altitude $z$ (Eq.~\ref{eq:cneps}). The sonic data give access to the sonic seeing $\epsilon_s$ defined { from Eq.~\ref{eq:cneps} by replacing the upper limit of the integral by any value $h$}
between $h_0$ and the height of the {highest} functionning sonic (ideally 45m when the corresponding anemometer gives valid data). The wavelength $\lambda$ was set to 500~nm. This quantity $\epsilon_s$ is the contribution of the surface layer (between $h_0$ and $h$) to the seeing. Computation of the integral was performed assuming a second order polynomial dependance of $\ln C_n^2(z)$ with the altitude. The three coefficients of the polynom were adjusted by least square on each set of simultaneous \cn\ values. For situations where only two anemometers were functionning, we computed a first-order polynomial fit of $\ln C_n^2(z)$ instead of a second order one.

Estimation of the sonic coherence time $\tau_s$ is very similar to the seeing, since the anemometers provide the wind speed and the constant \cn. 
{As for the seeing, it is defined from Eq.~\ref{eq:cntau0} by replacing the upper limit of the integral by $h$.  $\tau_s$ is } 
 the contribution of the atmospheric layer between $h_0$ and $h$ to the total coherence time $\tau_0$. The integral is computed the same way as the seeing. 

\subsection{Error analysis}
\subsubsection{Error on $C_T^2$}
\label{par:errct}
From Eq.~\ref{eq:calcct} one can estimate the standard deviation $\sigma_{C}$ of the constant $C_T^2$ averaged over a 1~mn data sample (which represents a number $n=600$ values at the sampling period $\delta t=0.1$~s). Assuming independence between consecutive sets of measurements we would have
\be
\frac{\sigma_{C}}{C_T^2}=\frac{1}{\sqrt{n}} \left[4\frac{\sigma^2_{(\delta T)}}{(\delta T)^2}\: + \: \frac{4}{9} \frac{\sigma^2_v}{v^2}
\right]^{1/2}
\ee
where $\sigma^2_v$ is the variance of $v$, $\delta T=[T(t)-T(t+\delta t)]$ and $\sigma^2_{(\delta T)}$ is its variance. However at our sampling rate there remains some correlation between consecutive measurements. Indeed the temporal autocorrelation function of the temperature shows a negative exponential decay at the origin, with a damping time of several tenths of seconds, sometimes 1~s. Taking 0.5~s for the correlation time of the temperature, the number $n$ of independant measurements in 1~minute is $n=120$ We will assume $n=100$ in the following.

The lowest value of $\delta T$ is the precision of the temperature fluctuation measurement, i.e. $\delta T=0.1$~K (Table~\ref{tab:sonics}). Its variance is $\sigma^2_{(\delta T)}=2 \sigma_T^2-2\, \mbox{Cov}[T(r),T(r+\delta t)]$, $\sigma_T^2$ being the variance of the temperature, and Cov$[T(r),T(r+\delta t)]$ the covariance between successive temperature measurements. A conservative value of $\sigma^2_{(\delta T)}$ is then $2 \sigma_T^2$. Since the temperature is measured with an accuracy of 0.1~K, we have $\sigma^2_{(\delta T)}\simeq 0.02$~K$^2$ which was verified experimentally on the data. The first term of the sum in the above equation is then of the order of $4\frac{\sigma^2_{(\delta T)}}{(\delta T)^2} \simeq 8$.

The term $\sigma_v$ contains both the accuracy of the anemometers for wind speed measurement (0.03 m/s, see Table~\ref{tab:sonics}) and the fluctuations of the wind speed during a 1~mn time interval. It was estimated experimentally to $\sigma_v\simeq 0.5$~m/s. Taking a small value of $v=1$~m/s, which is around the 10~\% percentile of the wind speed distribution (\cite{2008SPIE.7012E.147T}, the term $\frac{4}{9} \frac{\sigma^2_v}{v^2}$ is less than 0.1 and thus negligible in the above sum.

Finally the relative error on the structure constant $C_T^2$ (assuming $n=100$) is $\frac{\sigma_{C}}{C_T^2}\simeq 28\%$ (it would have been 12\% with $n=600$).

\subsubsection{Error on $C_n^2$}
From Eq. \ref{eq:ctcn} the error $\sigma_n$ on the refractive index structure constant can be expressed as
\be
\frac{\sigma_n}{C_n^2}=\left[\left(\frac{\sigma_{C}}{C_T^2}\right)^2+4 \left(\frac{\sigma_{P}}{P}\right)^2+16 \left(\frac{\sigma_{T}}{T}\right)^2
\right]^{1/2}
\ee
with $\frac{\sigma_{P}}{P}$, the relative error on the pressure, being of the order of $2\%$  (REF) and $\frac{\sigma_{T}}{T}\simeq 0.05\%$ this error budget is largely dominated by the error on $C_T^2$. Therefore the relative error on \cn\ is also of the order of 28\%.
\subsubsection{Error on the seeing}
The sonic seeing $\epsilon_s$ is defined by Eq. \ref{eq:sonicseeing}. A crude approximation of the integral is given by
\be
S=\int_{h_0}^{h} C_n^2(z)\, dz \; \simeq \; \sum_{i=1}^5 C_{ni}^2 \Delta z_i 
\ee
 where $C_{ni}^2$ is the structure constant given by the sonic number $i$, and $\Delta z_i$ the distance between two consecutive sonics ($\Delta z_i\simeq 7$m is almost the same for each sonic pair). The error $\sigma_S$ of this integral is deduced from the individual errors $\sigma^2_{n}$ of the \cn\ of each sonic. We have, assuming negiligible uncertainty for $\Delta z_i$
\be
\sigma_S=\left[\sum  \sigma^2_{ni} \Delta z_i^2 \right]^{1/2}\simeq 0.28\, \left[\sum C^4_{ni} \Delta z_i^2 \right]^{1/2}
\ee
the factor 0.28 above comes from the fact that the relative error on \cn\ is about 28\%. Therefore the relative error on the sonic seeing $\epsilon_s$ is
\be
\frac{\sigma_{\epsilon s}}{\epsilon_s}=\frac{3}{5} \frac{\sigma_S}{S} =0.17 \frac{\left[\sum C^4_{ni} \Delta z_i^2 \right]^{1/2}}{\sum C_{ni}^2 \Delta z_i} 
\ee
This expression depends on the geometry of the \cn\ profile. It would equals $0.17/\sqrt{5}\simeq 8\%$ for a flat profile, and is of the same order of magnitude for {the actual measured profile}.
%

\subsection{Bias due to spatial filtering}
\label{par:bias}
{As it will be seen later in Section~\ref{par:results}, } sonic estimations for \cn\ seem to be smaller than radiosoundings ones at the lowest altitudes ($h=8$m, $h=17$m, $h=23$m). Indeed sonic anemometers are known to introduce different bias effects (\cite{2015JPhCS.595a2036T}, in particular a spatial average over the sonic impulse path. The measurement of temperature with a sonic is made between the pair of ultrasonic transducers along the vertical direction $z$. The distance between these transducers is 15~cm: small spatial fluctuations of the temperature field in the $z$ direction are smoothed over twice this distance i.e. $l'=30$cm, because the impulse make a round trip between the sensors. An additionnal filtering occurs in the horizontal direction: as the integration time is not zero ($\delta t=0.1$), sonic impulses cross a slice of atmophere of length $v\delta t$ in the wind direction (mainly horizontal, the vertical component is weak). The effect of these two spatial filterings on the constant \ct\ can be derived by writing the  structure functions $D_T^2(\rho)$ as 
\be
D_T^2(\rho)=2\; (\sigma_T^2-\Gamma_T(\rho))
\ee
where $\sigma_T^2$ is the variance of temperature fluctuations, and $\Gamma_T(\rho)$ the cross-correlation between temperatures measured at times $t$ and $t+\Delta t$ (as discussed in \S~\ref{par:cnsonic}). Both quantities can be written as 3D Fourier integral over the spatial frequencies $(f_x,f_y,f_z)$. In the case of the variance we have
\be
\sigma_T^2 \; = \; \int\!\!\!\int\!\!\!\int_{-\infty}^{\infty} W(f_x,f_y,f_z)\, d^3 f
\ee

$W(f_x,f_y,f_z)$ the 3D power spectrum of the temperature, is proportionnal to $|f|^{-\frac{11}{3}}$ in the range $1/L_0 < |f| < 1/l_0$ (\cite{1971etaw.book.....T}, 
$l_0$ and $L_0$ being the inner and outer scales of turbulence. Neglecting very low frequencies and assuming $l_0=0$ we have
\be
\sigma_T^2 \; \propto \; \int\!\!\!\int\!\!\!\int_{\frac 1{L_0}}^{\infty} |f|^{-\frac{11}{3}}\, d^3 f
\ee
The vertical spatial  filtering of $\sigma_T^2$  is expressed by introducing the cutoff frequency  $\frac 1{l'}$ on the integral on $f_z$. Similarly, a cutoff frequency  $\frac 1{v \delta t}$ is introduced in one of the two horizontal variables (we chose $y$, but $x$ would give the same result).
\be
\sigma_T^{'2} \; \propto \; \int_{\frac 1 {L_0}}^{\infty}\!\!\!\int_{\frac 1 {L_0}}^{\frac1{v \delta t}} df_x df_y \int_{\frac 1{L_0}}^{\frac{1}{l'}} |f|^{-\frac{11}{3}}\, d f_z
\label{eq:ratioct2}
\ee
so the measured variance $\sigma_T^{'2}$ would be lower than the expected one $\sigma_T^2$. The ratio $\sigma_T^{'2}/\sigma_T^2$ depends on $L_0$ and $v$; the integration of eq.~\ref{eq:ratioct2} gives 0.8 for $L_0=15$m and $v=6$m/s.  This bias on the variance propagates on the structure function $D_T^2(\rho)$ and therefore on the constants \ct\ and \cn, which are then underestimated. We define as ``bias'' the ratio of measured to expected value of \ct .

We could compensate the individual measurements from this bias, using the instantaneous wind speed provided by the sonics, and the linear model for $L_0$ described in \S \ref{par:L0}. Median values found for the biases are reported in table~\ref{table:biaisct2} (the error bar on these median bias was found of about 10\%). It appears that the lowest sonics are significantly affected (underestimation by a factor ~35\% for the 8m sonic).

\begin{table*}
\begin{center}
 \begin{tabular}{l|rrrrrr}
 Sonic height    $[m]$                 &  8    &  17   & 23   & 31   & 39   &  45 \\ \hline
 $L_0 [m]$                     &    7  &  11   &  12  &  16  & 19  &  21 \\ 
 Bias on \ct\  (ratio measured/expected)    & 0.65  & 0.69 & 0.75 & 0.77 & 0.81 & 0.81\\ 
 \end{tabular}
 \end{center}
\caption{Estimation of the mutiplicative bias on the constant \ct\ measured by the 6 sonics. The uncertainty is around 10~\%.}
\label{table:biaisct2}
\end{table*}

\subsection{Comparison of the sonic \ct\ data with microthermal measurements}

\begin{figure}
\includegraphics[width=8cm]{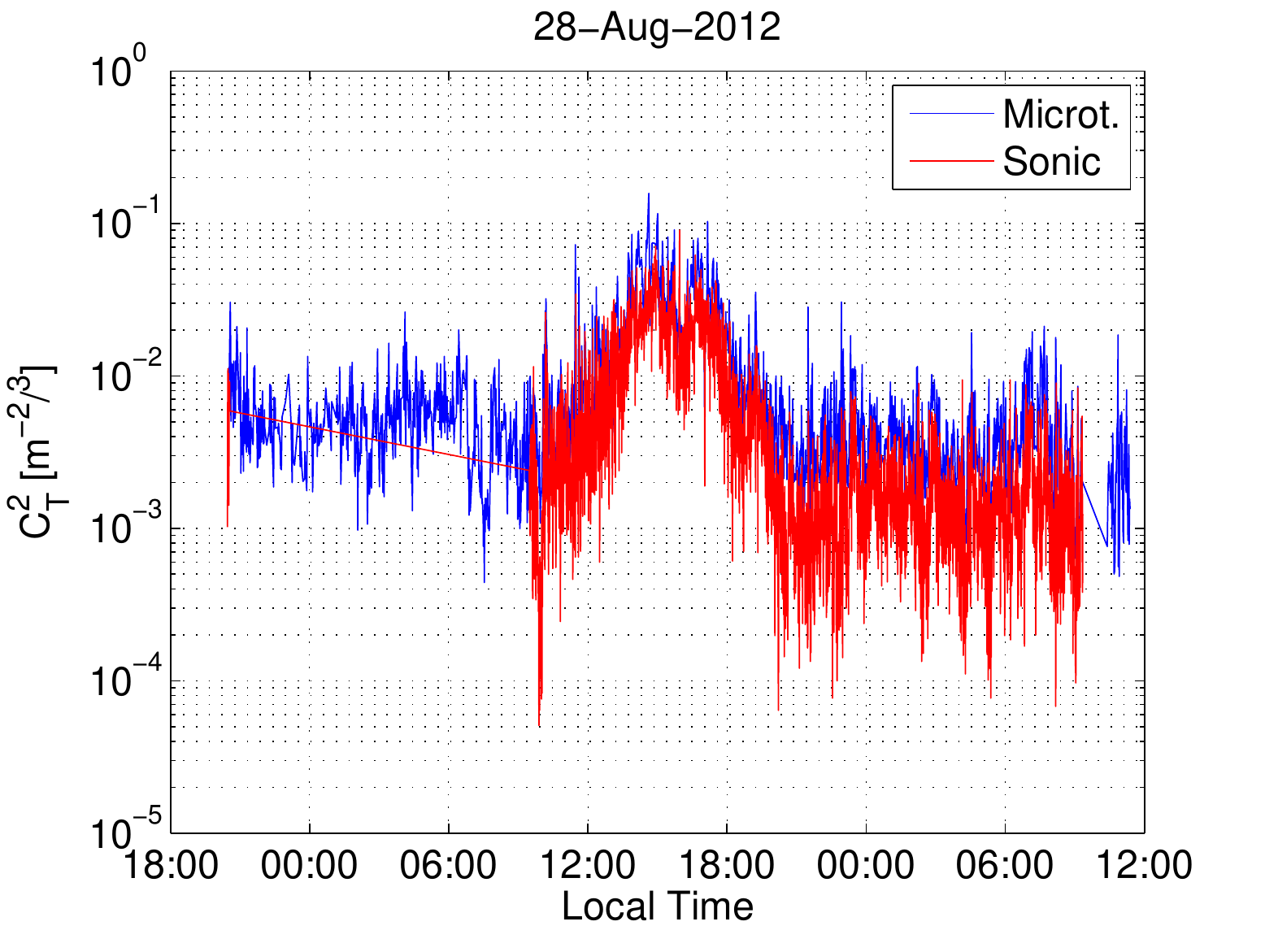} \ 
\includegraphics[width=8cm]{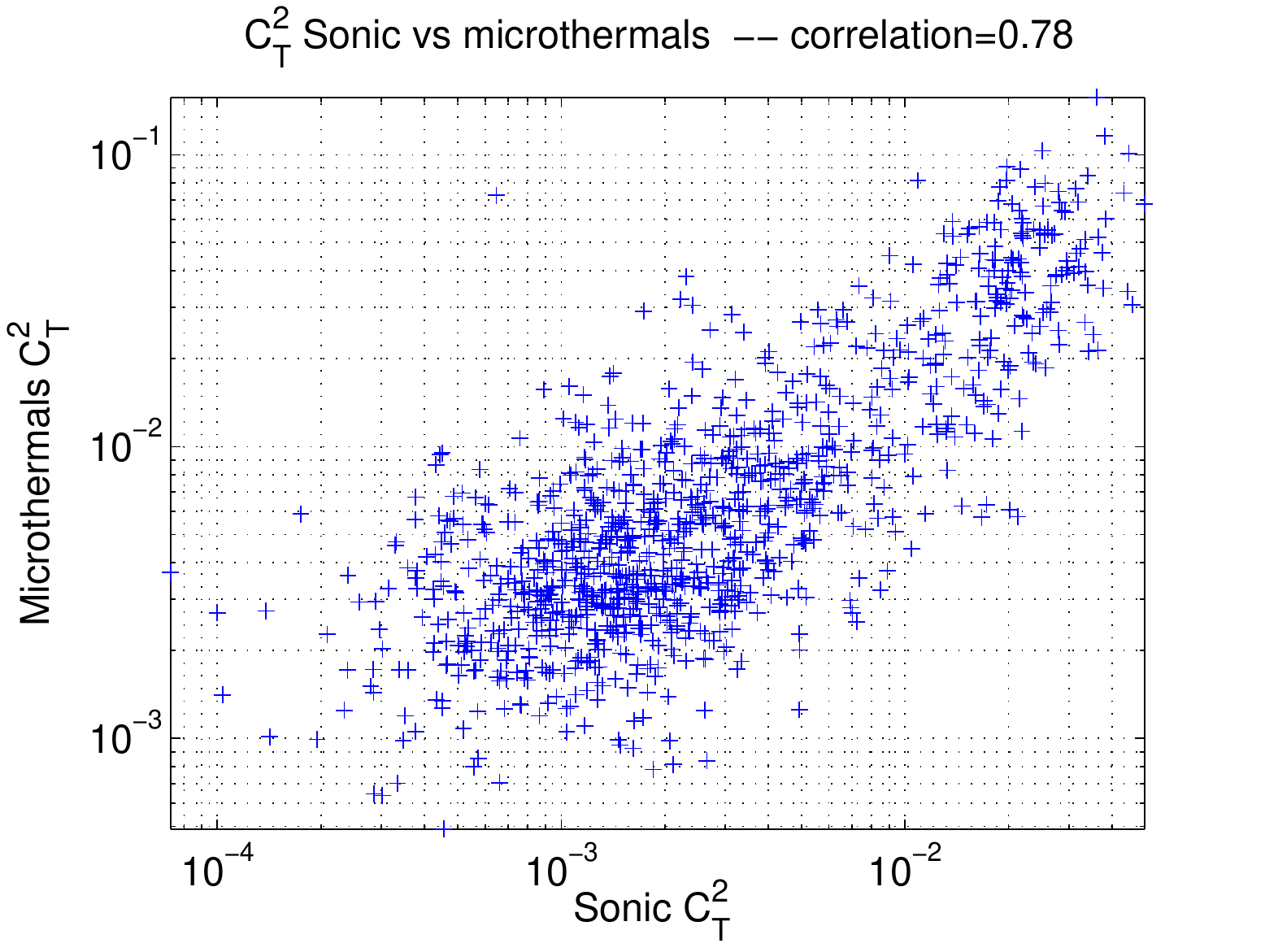} 
\caption{Simultaneous measurements of \ct\ at Nice observatory in August 2012. Left: times series. Right: plot of sonic \ct\ versus microthermal \ct within coincidence intervals of 30 seconds.}
\label{fig:calibsonicmat}
\end{figure}

In order to check the reliability of \ct\ values estimated by the sonics, we made a series of simultaneous measurements with a set of two microthermal pairs (\cite{2005PASP..117..536A} fixed to a balsa staff near the sonic anemometer. This calibration was made at Nice Observatory in August 2012: we just had a sonic anemometer repaired after some seasons in Antarctica, and could make the measurements before shipping it back to Dome C. The two instruments ran simultaneously for a period of $\simeq$36 hours. Times series of the temperature structure function \ct\ are presented in figure~\ref{fig:calibsonicmat}. { These data were not compensated from spatial filtering as presented in previous section, we indeed expect the multiplicative bias to be around 0.8, given the wind speed and outer scale at the time of the observations}. The 12h gap in the data is due to a problem with the acquisition PC. 
{The two curves are in good agreement}. In particular they show a peak in mid-afternoon due to strong local turbulence caused by the heating of the building by the Sun. { There remains a factor 2.3 of difference in the values which would not be totally explained by the spatial filtering bias. We suspect a bad calibration of the microthermals.} Fig.~\ref{fig:calibsonicmat} displays also a plot of sonic \ct\ versus microthermal \ct, considering coincidence intervals of 30~s between the two instruments. The correlation coefficient is 0.78. Is was unfortunately not possible to perform such a calibration at Dome C during Antarctica winter conditions.

\section{Results}
\label{par:results}

\subsection{Large scale structure functions}
\begin{figure}
\includegraphics[width=8cm]{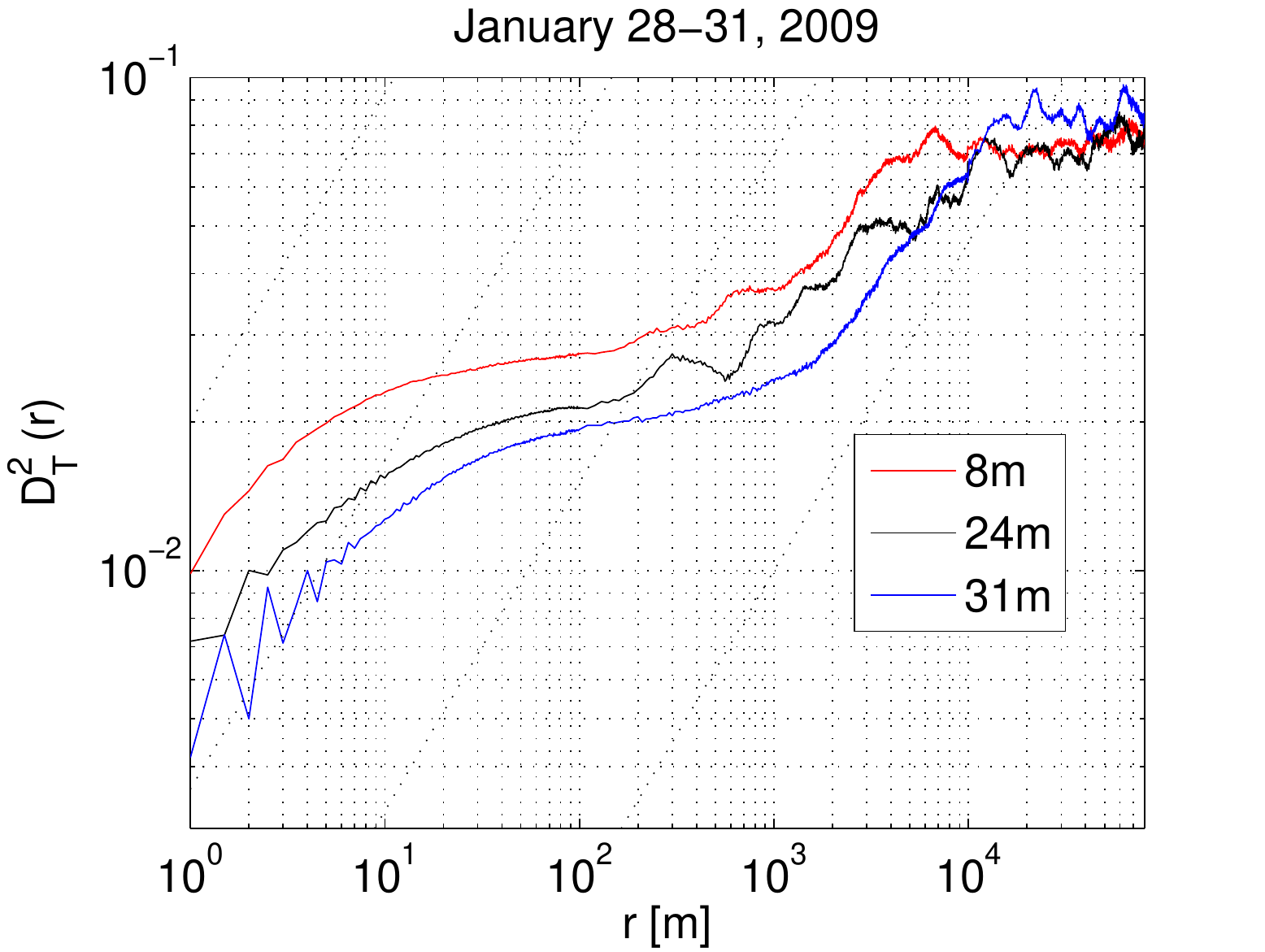}
\caption{Large scale structure function calculated on 4 days of data from Jan. 28th to Jan. 31st, 2009. Oblique dashed lines have a slope 2/3.}
\label{fig:fslong4j_20090128}
\end{figure}
(\cite{1991ApOpt..30..118C} reported observations of large-scale correlation between phase fluctuations of the light propagating through the turbulent atmosphere, and therefore large-scale correlation between thermodynamic parameters such as temperature or humidity. In their paper they reviewed different kind of such observations, and proposed an explanation based on the fact that at large scales (several km) the atmosphere is no longer isotropic and the troposphere may be considered as a ``thin layer'' in which a 2D turbulence structure develops. The consequence for the structure fonction of the temperature $D_T(\rho)$ is that it behaves as $\rho^{2/3}$ for $\rho<L_0$ (inertial zone), then a saturation regime is observed for $L_0 < \rho < L_S$ with $L_S\simeq$~1km. Then once again an inertial zone where $D_T(\rho)\propto \rho^{2/3}$ is observed for  $L_S < \rho < L_M$ with $L_M$~20km. Saturation occurs for $\rho>L_M$.

Typical wind speeds at Dome C are a few m/s near the ground, and the distance travelled by the air in one hour is about 10~km. It is then theoretically possible, with a sonic anemometer observing during one entire day, to probe structure functions $D_T(\rho)$ up to $\rho\simeq$~200km, and even more if we take continuous data over several days. Figure~\ref{fig:fslong4j_20090128} shows an example of large scale temperature structure function computed over 4~days of data, from January 28th to January 31st, 2009. Data for 3 sonics are displayed. The three curves show exactly the behaviour predicted by (\cite{1991ApOpt..30..118C}: a second inertial zone starting at $L_S=987$m for the sonic at $h=8$m, $L_S=727$m for $h=24$m and $L_S=1489$m for $h=31$m. Statistical significance becomes poor at larges $\rho$ and this is the reason why we considered a 4-days long data sample.

\begin{figure}
\includegraphics[width=8cm]{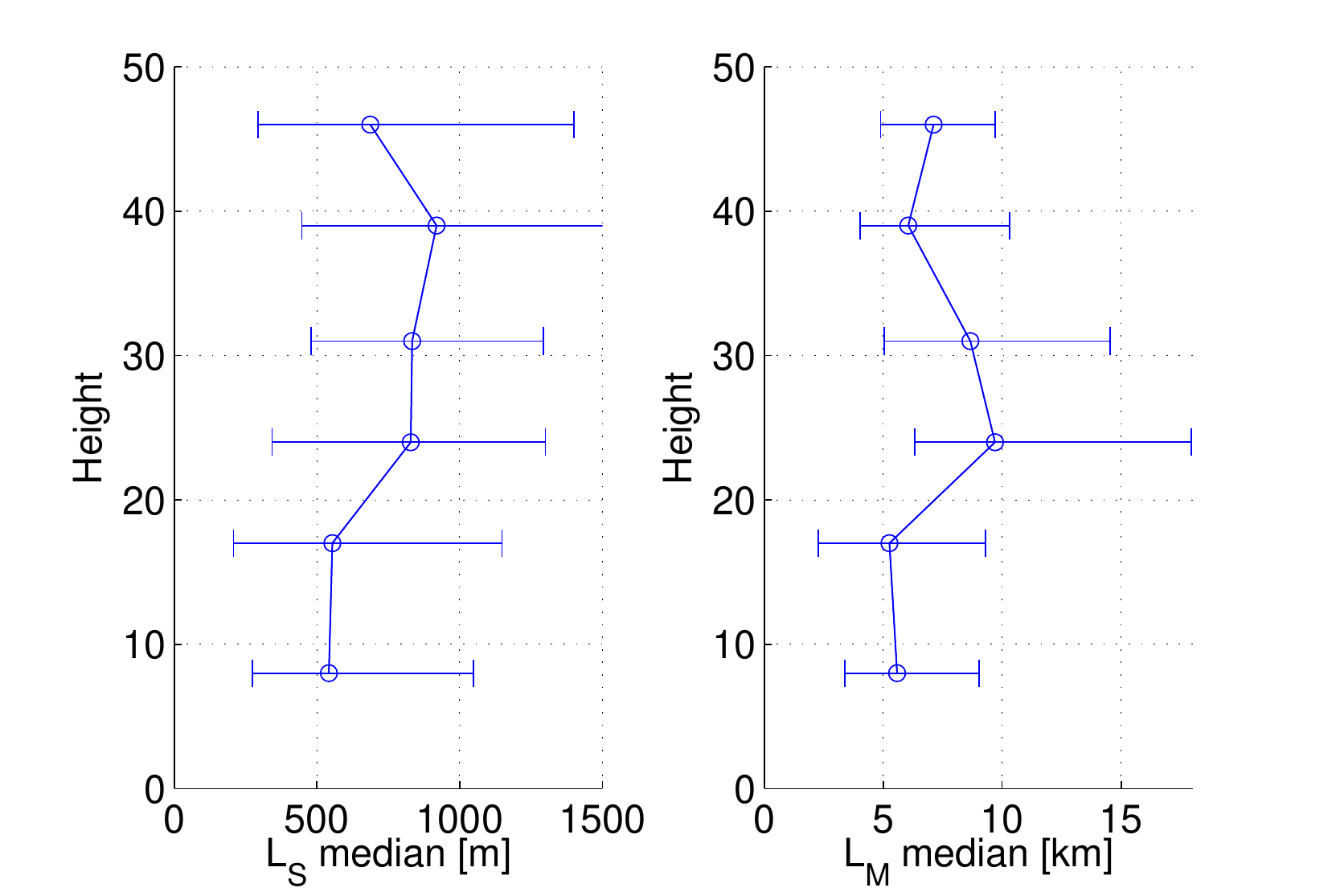}
\caption{Vertical profiles of $L_S$ and $L_M$ computed from 4-days structure functions for each sonic. Error bars represent the 50\% confidence interval.}
\label{fig:ls_lm_vs_h}
\end{figure}

From the whole set of sonic data, we could compute, for each sonic, 378 large structure functions spanning over 4 days. Figure~\ref{fig:ls_lm_vs_h} displays the variation of scales $L_S$ and $L_M$ as a function of the altitude, and show no significative dependence, as one could expect regarding to the large numbers involved. Similarly, we found no seasonal dependence of the scales. Global statistics for $L_S$ and $L_M$ are given in Table~\ref{table:ls_lm_stat}.

\begin{table}
\begin{tabular}{l|ccc}
     & median & 1st quartile & 3rd quartile\\ \hline
$L_S$ &  713.5 m & 317 m  & 1282 m\\
$L_M$ &  6.5 km  & 4 km  & 11 km  \\ \hline
\end{tabular}
\caption{Statistics of the scales $L_S$ and $L_M$ (boundaries of the second intertial zone) computed from 378 outer scales sapnning over 4 days data sets.
}
\label{table:ls_lm_stat}
\end{table}

\subsection{Statistics of \cn}
A total amount of 634000 values of valid estimations were obtained from the six sonic anemometers. Table~\ref{tab:statcn} presents the statistics of the structure constant \cn\ obtained at every height (the median value, and the 50\% confidence interval between the 25\% and the 75\% percentiles). At Dome C the conditions of turbulence vary strongly with the season (\cite{2009AA...499..955A} and we split the data into four samples corresponding to four seasons. We call here summer the period when the Sun is almost circumpolar (November -- January), winter when it never rises (May -- July), autumn and spring being the 3-months interseasons. Statistics for both seasons are also given in the table~\ref{tab:statcn}. About 50\% of the data were collected in summmer, 27\% in autumn, 12\% in spring and 11\% in winter.

\begin{figure}
\includegraphics[width=8cm]{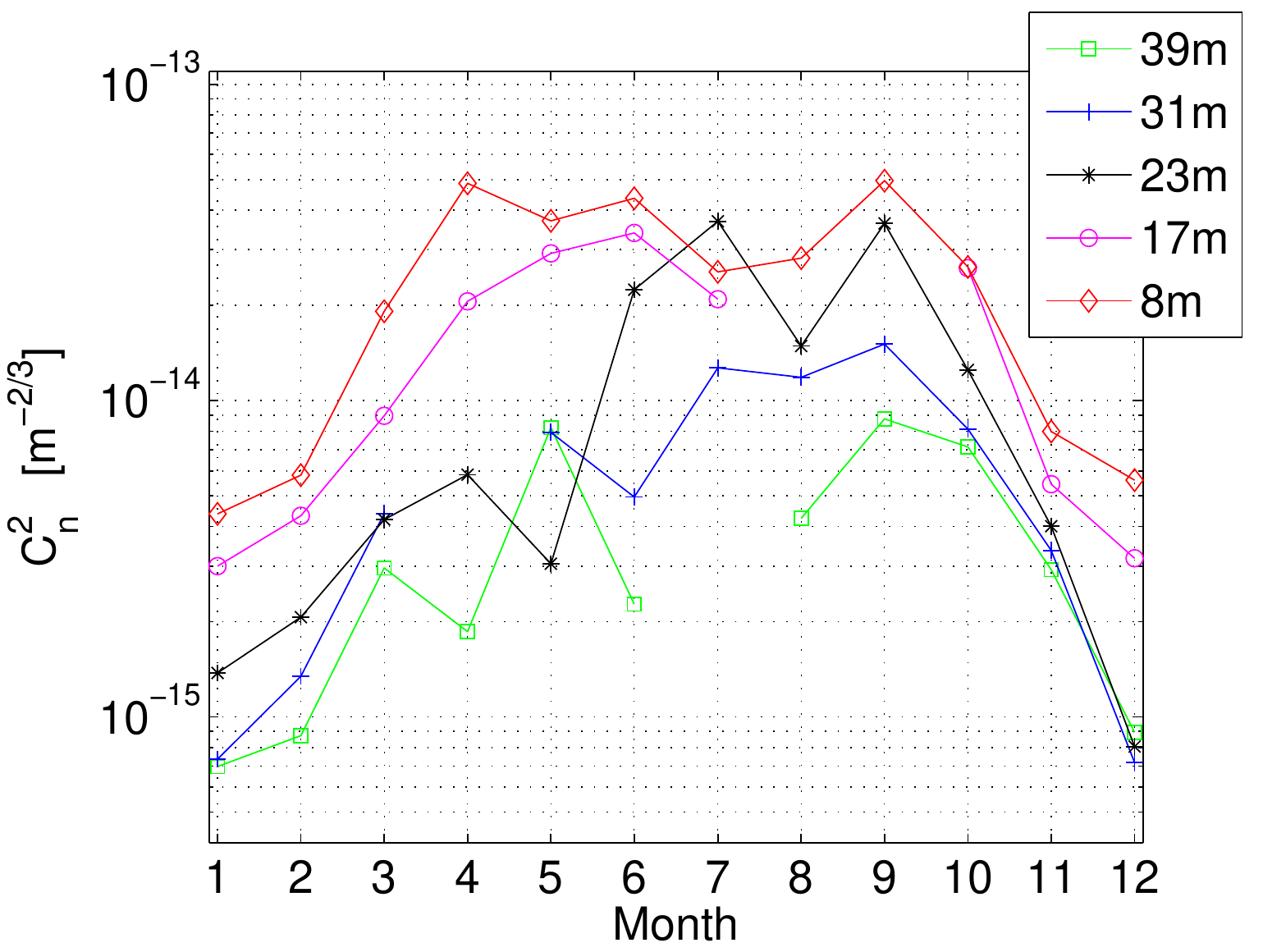}
\caption{Monthly median values of the structure constant \cn\ for different heights.}
\label{fig:cn2month}
\end{figure}

Our values of \cn\ confirm that the optical turbulence near the surface is the strongest in winter, \cn\ is 5 to 10 times higher than in summer. 
Dependance of \cn\ with the month is shown in Fig.~\ref{fig:cn2month} and display a sinusoid-like arch with a maximum in June-July for all the sonics. These are the same sinuso\"{\i}dal plots that were observed for the DIMM seeing times series (see Fig.~1 of (\cite{2009AA...499..955A}).

We compared our results with the values obtained by in-situ microthermal radiosoundings performed in 2005 during the first winterover (\cite{2008PASP..120..203T}. For this purpose we selected data from the period April -- October corresponding to the months where 34 balloon were launched. Fig.~\ref{fig:cn2prof} shows the median \cn\ profile from the sonic anemometers, the superimposed dashed line is the median values measured by the radiosoundings. We also plotted the data compensated from the bias due to spatial filtering. The sonic point at $h=45$m is not significative (too few data).
The profiles are comparable, balloon data fit with sonic error bars, though it seems that sonic points are still underestimated, even after compensation for the bias. Both curves show a rapid decay of \cn\ between 8~m and 45m. However the balloon profile presents a steeper slope: between 8~m and 40~m the \cn\ is divided by about 50 for the balloon data, and by 15 for the sonic data (20 if we take the bias-compensated values for sonic \cn).

\begin{figure*}
\includegraphics[width=15cm]{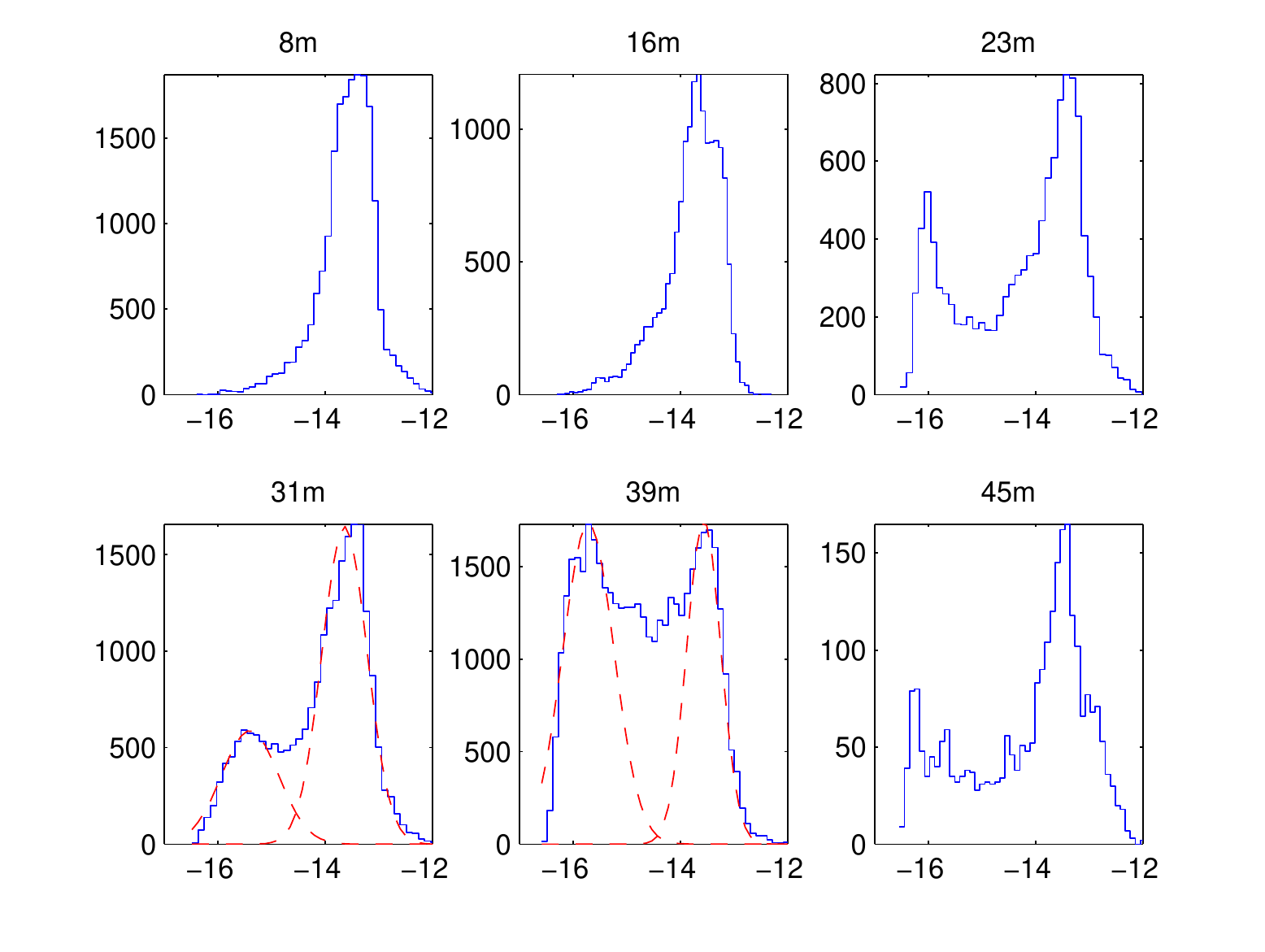}
\caption{Histograms of log \cn\ at different heights for the period April-September. Data corresponding to heights 31m and 39m were fitted by a sum of two Gaussian (dashed red curves).}
\label{fig:histo6cnhiv}
\end{figure*}

\begin{figure*}
\includegraphics[width=\textwidth]{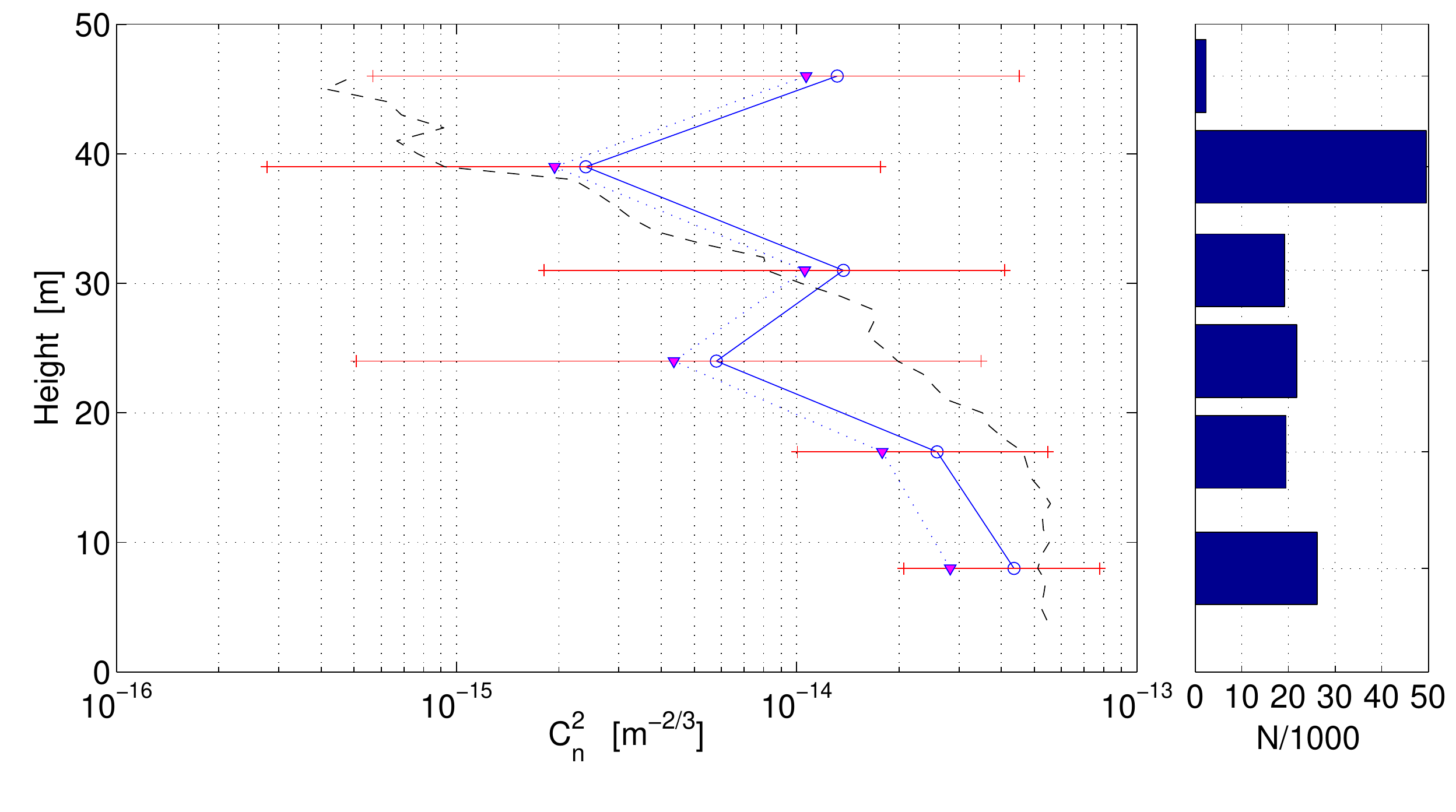}
\caption{\cn\ median profile for the period April-October. Blue circles: values corrected from the bias as described in \S\ref{par:bias}. Triangles : raw values. Error bars are the 50\% confidence interval around the blue circles. The dashed line is the median profile obtained in 2005 (March, 15th to October, 19th) from in-situ radiosoundings. The histogram on the right is the number $N$ of valid sonic data at each altitude.}
\label{fig:cn2prof}
\end{figure*}

Another interesting result is shown in figure~\ref{fig:histo6cnhiv}. It displays the 6 histograms of \cn\ measurements at each elevation for the period April to September, from 2008 to 2011. They show a bimodal distribution centered at values $C_n^2=2.\, 10^{-16}$~m$^{-2/3}$ and $C_n^2=2.\, 10^{-14}$~m$^{-2/3}$. The two bumps are very clear for upper heights (23m to 45m) but the left one (centered at $C_n^2=2.\, 10^{-16}$~m$^{-2/3}$) exists also at $h=$16m (with very small amplitude). These distributions are similar to DIMM histograms described in (\cite{2009AA...499..955A} and the interpretation is similar: the  bump on the right (resp. left) corresponds to situations where the sonic is above (resp. inside) the surface layer. There exist also a few intermediate situations corresponding to the cases where either the SL is not unique, but contains a second layer above, or the SL upper limit is just in front of the sonic and moves slightly up and down during the 1mn time interval other which the constant \cn\ is estimated. These situations correspond to 14\% of the measurements at $h=31$m and 22\% at $h=39$m.

From these histograms, it can be seen that the sonic at $h=$31m spends more time inside the surface layer (the right bump is higher than the left one). The situations are more equilibrated at $h=39$m. On these two particular histograms, we performed a least-square fit of the sum of two Gaussians (indeed log-normal functions since the abcissa is in log scale). The fits are drawn in dashed red on the curves. The ratio of the surface of the two Gaussians (left/right) is $r_{31}=0.44$ at $h=$31m and  $r_{39}=1.45$ at $h=39$m. A mere interpolation would give a ratio $r_H=1$ at $H=35$m, which is an estimation of the thickness of the surface layer. It is compatible with previous measurements as well as turbulence modelling (see (\cite{2012sf2a.conf..697A} and references therein).

\begin{table*}
\begin{center}
%
\begin{tabular}{c|c|c|c|c|c|}\hline
Height & \cn\ (total)          &  \cn\ (summer)         &  \cn\ (autumn)          & \cn\ (winter)         & \cn\ (spring)\\
$[m]$  & $[10^{-15} m^{-2/3}]$ & $[10^{-15} m^{-2/3}]$  & $[10^{-15} m^{-2/3}]$   & $[10^{-15} m^{-2/3}]$ & $[10^{-15} m^{-2/3}]$\\ \hline
8      & $8.7 \left[\begin{array}{c}3\\27\end{array}\right.$& $5.1 \left[\begin{array}{c}2\\14\end{array}\right.$   & $9.4 \left[\begin{array}{c}3\\29\end{array}\right.$    & $36.4 \left[\begin{array}{c}19\\66\end{array}\right.$       &  $30.2 \left[\begin{array}{c}11\\64\end{array}\right.$            \\\hline
17     & $6.2 \left[\begin{array}{c}2\\20\end{array}\right.$& $3.4 \left[\begin{array}{c}1\\10\end{array}\right.$  & $6.8 \left[\begin{array}{c}2\\21\end{array}\right.$    &  $27.9 \left[\begin{array}{c}12\\56\end{array}\right.$        &  $29.3 \left[\begin{array}{c}6\\128\end{array}\right.\,^{(*)}$            \\ \hline
24     & $2.5 \left[\begin{array}{c}0.5\\10\end{array}\right.$& $1.5 \left[\begin{array}{c}0.4\\6\end{array}\right.$ & $2.8 \left[\begin{array}{c}0.5\\10\end{array}\right.$  &  $3.1 \left[\begin{array}{c}0.4\\8\end{array}\right. \,^{(*)}$  &  $16.1 \left[\begin{array}{c}4\\39\end{array}\right.$            \\ \hline
31     & $2.2 \left[\begin{array}{c}0.4\\11\end{array}\right.$& $1 \left[\begin{array}{c}0.3\\4\end{array}\right.$ & $2.8 \left[\begin{array}{c}0.45\\13\end{array}\right.$   & $9.7 \left[\begin{array}{c}1.2\\25\end{array}\right.$       &  $9.8 \left[\begin{array}{c}1\\29\end{array}\right.$            \\ \hline
39     & $1.6 \left[\begin{array}{c}0.3\\9\end{array}\right.$& $1 \left[\begin{array}{c}0.3\\4\end{array}\right.$ & $1.5 \left[\begin{array}{c}0.2\\8\end{array}\right.$    & $2 \left[\begin{array}{c}0.3\\14\end{array}\right.$       &  $6.6 \left[\begin{array}{c}0.6\\25\end{array}\right.$            \\ \hline
45     & $0.9 \left[\begin{array}{c}0.2\\6\end{array}\right.$& $0.7 \left[\begin{array}{c}0.2\\4\end{array}\right.$ & $1 \left[\begin{array}{c}0.2\\8\end{array}\right.$   & $(*)$  &  $3.4 \left[\begin{array}{c}0.2\\22\end{array}\right.$            \\ \hline
\end{tabular}

\end{center}
\caption{Statistics of the structure constant \cn\ at different heights above the ground. Median value and percentiles 25\% and 75\% are given. The definition of the four seasons (summer, autumn, winter and spring) is given in the text. The symbol (*) indicates a low number of data for this height and this period.}
\label{tab:statcn}
\end{table*}

\subsection{Seeing statistics}
\begin{figure}
\includegraphics[width=8cm]{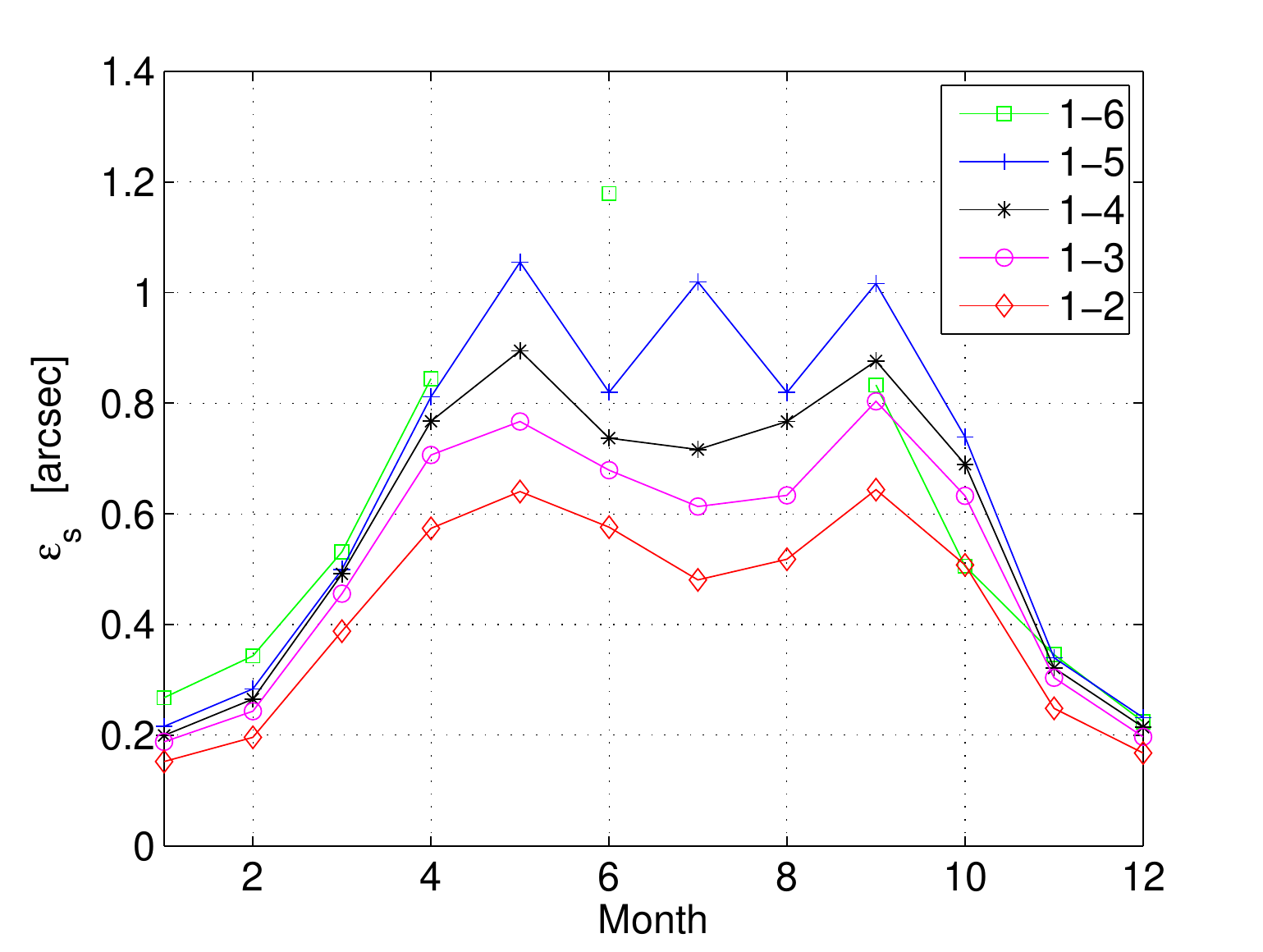}
\caption{Monthly median values of the surface layer seeing $\epsilon_s$ integrated between 8~m and the altitude of each sonic. Label such as ``1--3'' means an integration between the first sonic (h=8~m) and the third (h=24~m).}
\label{fig:seeingvsmonth}
\end{figure}

Using eq. \ref{eq:sonicseeing} we could calculate the seeing $\epsilon_s$ in the surface layer, between the first sonic at $h=8$m and all the other sonics, up to $h=45$m. The data sample used for computing the seeing represents about 20\% of the number used for \cn\ statistics. The reason for that is that we rejected the situations where the lower sonic either did not work or was the only sonic to work.  Figure~\ref{fig:seeingvsmonth} shows the montly evolution of the median of $\epsilon_s$. Error bars (50\% confidence intervals) are $\pm 0.1$~arcsec in summer and $\pm 0.4$~arcsec in winter. The curves have a sinusoidal shape very simular to the DIMM curves displayed in Fig.~1 of (\cite{2009AA...499..955A}.

Looking at the height 39~m it can be seen that during the April--October period the median seeing inside the surface layer (between 8m and 39m) is around 1~arcsec, which is consistent with the value of 1.2~arcsec derived from the 2005 radiosoundings (\cite{2008PASP..120..203T}. 


\subsection{The peculiar summer situation}
\begin{figure}
\includegraphics[width=8cm]{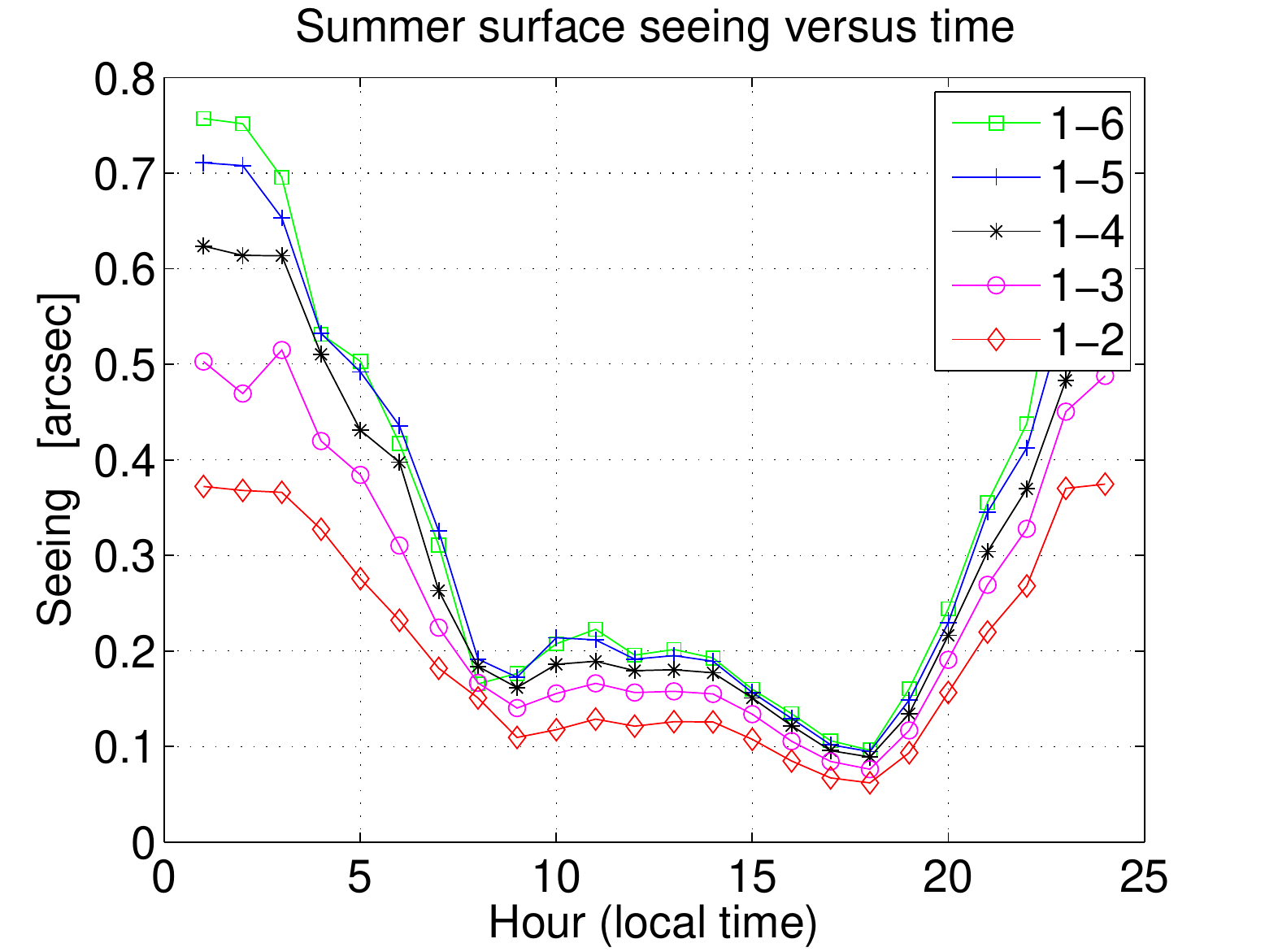}
\caption{Hourly median values of the surface layer seeing $\epsilon_s$ in summer (the period considered here is December 2008 where all 6 sonics were in operation), integrated between 8~m and the altitude of each sonic. Label such as ``1--3'' means an integration between the first sonic (h=8~m) and the third (h=24~m).}
\label{fig:seeing_bl_summer_vs_time}
\end{figure}

Turbulence monitoring at Dome C began in 2003 with the first summer site-testing campaigns. It was noticed (\cite{2005AA...444..651A}  that the seeing was very low with a deep minimum every day near 5pm local time. Indeed, meteo radiosoundings have shown that the vertical temperature profile is flat twice a day: in the morning around 10am local time, and in the middle of the afternoon (see fig. 10 of (\cite{2005AA...444..651A}). The installation of the sonics in 2007 allowed a monitoring of the surface layer seeing $\epsilon_s$ in summer, and particularly its dependence with time. The result, computed from December~2008 sonic data, is shown in fig.~\ref{fig:seeing_bl_summer_vs_time} and is very interesting. We found that $\epsilon_s$ also shows, as expected, a deep minimum near 5pm local time. At this minimum, the curves labelled ``1-4'', ``1-5'' and ``1-6'' are coincident: it means that the quasi totality of the surface layer turbulence is below the 4th sonic (height 31m). But there is also another minimum, though a little  less deep, at 9am. It corresponds to the temperature gradient inversion observed in the morning. And this secondary minimum was not present on the seeing curve provided by the DIMM during the ``night'' hours, i.e. from 11pm to 2am, (there is no night in summer, but a period when the Sun is low on the horizon), the surface layer seeing is rather strong, and spread over the whole height of the tower. The value of $\epsilon_s$ beween 8m and 45m at 0am is $\epsilon_s\simeq$~0.75~arsec and the total seeing $\epsilon_d$ observed by the DIMM at the same time is $\epsilon_d\simeq$~0.9~arsec; the difference $[\epsilon_d^{5/3}-\epsilon_s^{5/3}]^{3/5}$ gives 0.4~arcsec, which is almost exactly the free atmosphere seeing as mentionned by (\cite{2009AA...499..955A}.

\subsection{Coherence time}
\begin{figure}
\includegraphics[width=8cm]{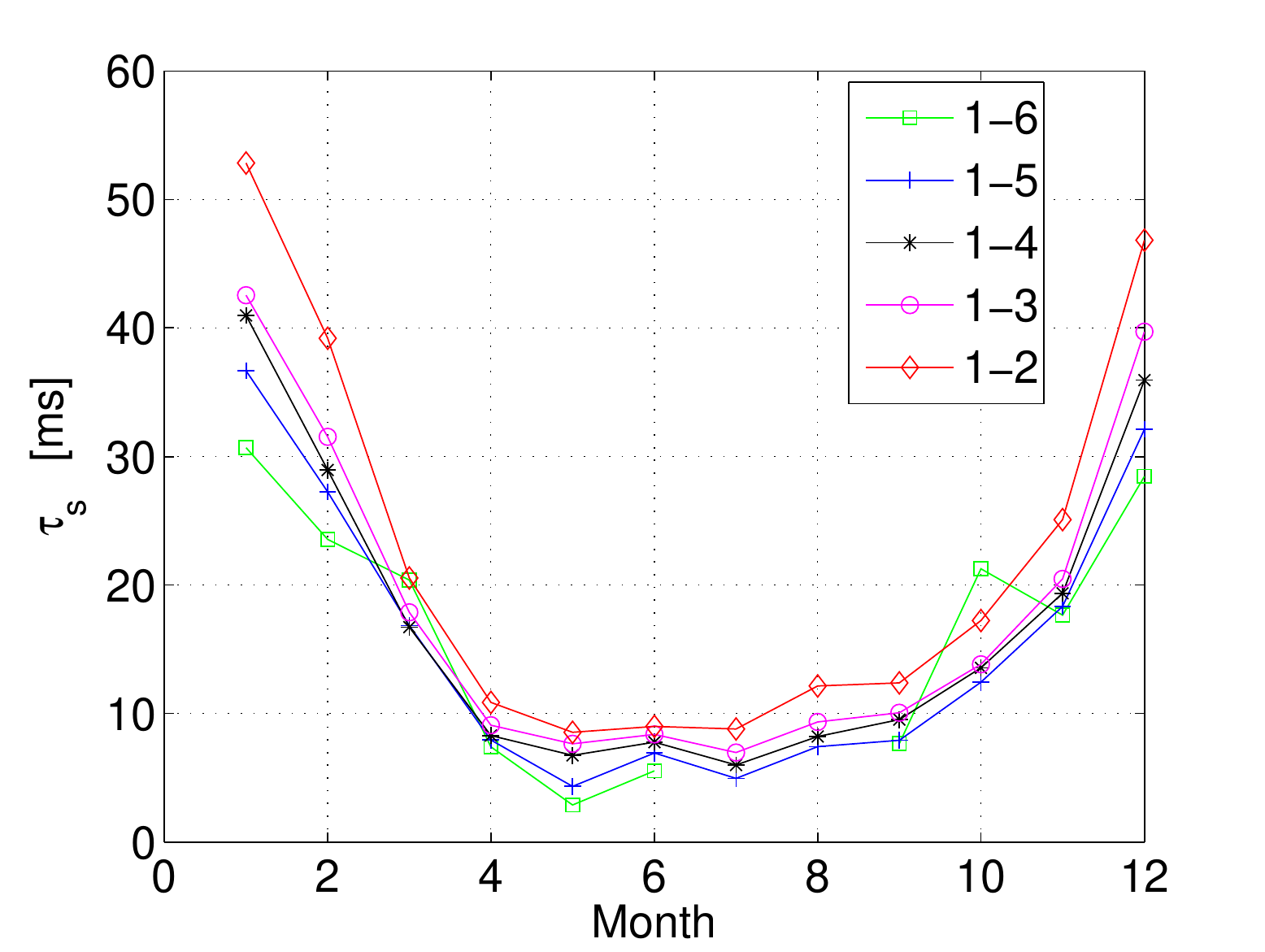}
\caption{Monthly median values of the coherence time surface layer $\tau_s$ integrated between 8~m and the altitude of each sonic.}
\label{fig:tau0vsmonth}
\end{figure}

Statistics of the sonic coherence time $\tau_s$ (i.e. contribution of the surface layer) were computed on a sample containing about $300\,000$ valid data. A strong dependence with the season was found as expected. 
Fig.~\ref{fig:tau0vsmonth} displays the evolution of the monthly median values of $\tau_s$ for each layer. Unsurprisingly the turbulence is faster in winter, and the curves display a sinuso\"{\i}dal arch which is minimum in July--August with a value $\tau_s\simeq 8$~ms. 

A partial comparison of our results can be made with values of (\cite{2008PASP..120..203T}. In their Table~2 they give the statistics of the coherence time $\tau_0$ computed at $h=8~m$ and $h=33~m$ in autumn, winter and spring. Making use of the integral expression of $\tau_0$ given in Eq.~\ref{eq:cntau0}, we could estimate the contribution $\tau_s^b$ of the layer between 8~m and 33~m, and compare it with our value of $\tau_s$ in the layer between the sonics 1 and 4 ($h=31$~m). Results are displayed in Table~\ref{table:tau0s_bal_sonic}, and show that the two instruments give consistent measurements.

\begin{table}
\begin{tabular}{l|cc}
       & $\tau_s^b$ (balloon) & $\tau_s$ (sonics) \\ \hline
Autumn & 18~ms & 24~ms [12 -- 45] \\
Winter & 7~ms  & 7~ms [5 -- 12] \\
Spring & 16~ms & 12~ms [7 -- 29] \\ \hline
\end{tabular}
\caption{Comparison of the surface layer coherence time computed from balloon data (Trinquet et al, 2008a) and sonic data. Balloon values are calculated for the layer between 8m and 33m. Sonic values are calculated for the layer between 8m and 31m: median value is given as well as the 50\% confidence interval between brackets.
}
\label{table:tau0s_bal_sonic}
\end{table}

\section{Conclusion}
We have presented measurements of optical turbulence inside the surface layer at Dome C. These measurements represent 6 years of data collected by up to 6 sonic anemometers placed on a 45m high tower. Sonic anemometers were chosen as an alternative to microthermal sensors, since these sensors are too fragile for the harsh conditions of Dome~C.

Operation of sonic anemometers appeared to be easy in Summer, with temperatures around -30$^\circ$C and the largest amount of good quality data was collected in this period. Things become more difficult with the drop of temperatures in April. Despite the heating resistances wrapped around the arms of the sonics, we cannot totally prevent the deposit of frost and a lot of data were unusable (for the period May-August, only 10\% of the data was useful). It was also necessary to climb periodically (about once a week) to the tower to remove the snow accumulated on the sonics. And a lot of technical problems were met as described in section~\ref{par:instrument}.

Despite these difficulties of operation, and thanks to the long running period, interesting results could be derived. Temperature structure functions are the basis of the work, and we found that they behave as predicted, with an inertial regime in $\rho^{2/3}$ and a saturation for larger scales of a few tens of meters. Dynamic outer scales could be measured at the intersections on these two regimes. They increase with altitude with a slope $\simeq 0.4$ as predicted for isotropic turbulence.
Structure functions were probed for large scales and show a second inertial zone in the range $\rho \in [1 - 10]$~km as predicted by (\cite{1991ApOpt..30..118C}. To our knowledge this is the first time that such observations are reported since the 1991 paper.

Temperature structure constants \ct\ were compared to simultaneaous microthermal measurements which gave satisfactory coherence (though the comparison was made in temperate temperature conditions). Some bias was observed between the sonic \cn\ and the balloon-borne microthermal radiosoundings, and could be explained by spatial filtering of temperature fluctuations {due} to the size of the sonic arms and the finite integration time.
The overall behaviour of the surface layer turbulence at Dome C is {consistent with} previous studies by different technologies: balloon-borne microthermals (\cite{2008PASP..120..203T}, DIMMs at different elevations (\cite{2009AA...499..955A} and SODAR (\cite{2014AA...568A..44P}. They are also consistent with turbulence modelling by (\cite{2011MNRAS.411..693L}. The sonics allowed a new estimation of the thickness of the surface layer in winter (35m) which agrees here again with previous estimations.

Integrated parameters (seeing and coherence time) could be calculated in the surface layer. The large amount of data allowed to make significant statistics and to probe their dependence with the season. In particular, this is the first time that these parameters are measured during the summer. We found that the surface layer is responsible of about 1~arcsec of the total seeing in winter. In summer, the surface layer seeing exhibits every day two minima with very low values, one in the afternoon and one in the morning, these two minima correspond to the inversion of the vertical temperature gradient observed by the radio-soundings. 

This study show that sonic anemometers are a viable option to undertake surface layer turbulence monitoring. However careful attention has to be paid to data processing, especially for bias compensation and bad points removal, and it was one of the difficulties of this work. Measurements obtained here are consistent with what was previouly know from various technologies and gives confidence on the whole results set. One of the advantages of the sonics {is} the possibility to infer the properties of temperature and refractive index structure functions at small and large spatial scales, which is not so common in the panorama of available instruments.  The weak point was the sensitivity to harsh climatic conditions, but it's a common problem for every instrument operating in  Antarctica. This study was conducted within the polar program AstroConcordia, which terminated in 2012. This paper is one of the last from our group, on the topic of Dome C site characterization for astronomical purposes.

\section*{Acknowledgements}
The authors gratefully acknowledge the polar agencies IPEV and ENEA for their logistic and financial support to our programs (IPEV/AstroConcordia, IPEV/CALVA and INSU/CLAPA). Thanks are also due to the US NSF and the french agencies INSU and ANR for funding. We are in debt to the Dome C local staff and winter-overs from 2007 to 2012 for their assistance, in particular to people who climbed the tower to remove the snow on the sonics. Thanks also to Herb Zimmerman from Applied Technologies for technical support and advices throughout these years.

\bibliographystyle{mnras}
\bibliography{aristidi} 

\end{document}